\documentclass[nofootinbib]{revtex4-2}
\usepackage{hyperref}
\hypersetup{colorlinks,allcolors=black}
 \usepackage{graphicx}
 \usepackage{amsmath}
 \usepackage{amssymb}
 \usepackage[utf8]{inputenc}
 \usepackage{natbib}
 \usepackage{slashed}
 \usepackage{subfigure}
\usepackage{pifont}

\begin{document}
\title{A Multiplicative Formulation of the Higgs Lagrangian and the Fermion Mass Hierarchy between the charged leptons and heavy quarks}
\author{Suppanat Supanyo}
\affiliation{The Institute for Fundamental Study (IF), Naresuan University, Phitsanulok 65000, Thailand}
\author{Sikarin Yoo-Kong}
\affiliation{The Institute for Fundamental Study (IF), Naresuan University, Phitsanulok 65000, Thailand}
\author{Lunchakorn Tannukij}
 \affiliation{Department of Physics, School of Science, King Mongkut's Institute of Technology Ladkrabang, Bangkok 10520, Thailand}
\begin{abstract}
We propose a multiplicative formulation of the Higgs Lagrangian, derived from the inverse problem 
in the calculus of variations, as an alternative framework to investigate the fermion mass hierarchy. 
In this setup, fermion masses emerge as discrete quantities determined by a finite set of scaling factors, 
thereby allowing the observed charged-lepton and heavy-quark masses to be accommodated 
without introducing arbitrarily small parameters. 
In addition, this framework admits specific solutions where the Yukawa couplings of the charged leptons 
and heavy quarks converge to a universal value, approximately coinciding with the Higgs self-coupling. 
This numerical coincidence provides a potential hint of an underlying dynamical structure 
that correlates the Higgs sector with the fermion masses.
Furthermore, the background-dependent Higgs self-interactions are found to decrease asymptotically, ensuring perturbative consistency in the large-field regime and suggesting possible extensions toward ultraviolet completion.
\end{abstract}

\maketitle

\section{Introduction}

The origin of the fermion mass hierarchy remains a profound open question within the Standard Model (SM). The wide range of observed Yukawa couplings, spanning from $\mathcal{O}(1)$ to $\mathcal{O}(10^{-6})$, strongly suggests the presence of an underlying fundamental structure beyond the SM. Numerous theoretical frameworks have been proposed to address this hierarchy, including the introduction of new symmetries, effective field theories, and extra-dimensional constructions. Among the proposed extensions of the SM, the SO(10) grand unified theory (GUT) \cite{PhysRevD.50.3529, PhysRevD.24.1895, PhysRevD.27.600, PhysRevD.46.4004, PhysRevLett.79.4748, PhysRevD.61.035008, Djouadi:2021bwu, Djouadi:2022gws} stands out due to its renormalizability, its elegant unification of matter representations, and its predictive power regarding fermion quantum numbers and charge quantization. However, minimal implementations of SO(10) often predict mass relations such as $m_d = m_e$ and $m_s = m_\mu$ at the unification scale, which are incompatible with experimental data unless substantial corrections are included. Within the effective field theory framework, the Froggatt–Nielsen (FN) mechanism \cite{Froggatt:1978nt, PhysRevD.98.011701, Wang:2011ub} offers an appealing explanation of the Yukawa hierarchy via the introduction of horizontal U(1) flavor symmetries. Nonetheless, the predictivity of the FN framework is limited by the presence of undetermined $\mathcal{O}(1)$ coefficients, both in the choice of U(1) charge assignments and in the effective couplings. Due to present limitations in experimental sensitivity, it remains inconclusive which of the proposed models captures the true underlying mechanism of flavor hierarchy. The question of fermion mass hierarchy thus remains open to further exploration. 

In this work, we propose an alternative framework inspired by the inverse problem in the calculus of variations \cite{douglas1941solution, hojman1981inverse, sarlet1982helmholtz, Henneaux_1984}. This formalism allows for the construction of Lorentz invariant nonstandard Lagrangians beyond the conventional additive form $\mathcal{L} = T - V$. In particular, we explore a class of multiplicative scalar field Lagrangians of the form
\begin{align}
\mathcal{L}=F(\partial_\mu\phi^\dagger\partial^\mu\phi)\, f(\phi^\dagger\phi),
\end{align}
which also appear in contexts of string inspire model such as k-essence \cite{PhysRevD.63.103510, PhysRevD.105.064058, PhysRevD.103.043518}, Dirac–Born–Infeld (DBI) models \cite{PhysRevD.83.101301, PhysRevD.70.123505, PhysRevD.77.023511}.
Here, $F$ and $f$ were originally derived in \cite{PhysRevD.106.035020} and will be reintroduced in next section. 
Within this framework, we demonstrate that if the Higgs sector of the electroweak theory (EW) is in the summation overall of the multiplicative Lagrangian form, the resulting non-canonical kinetic structure can induce a finite and discrete fermion mass spectrum. Moreover, this structure can provide a hierarchical pattern of fermion mass across generations of the charged lepton and heavy quarks. Notably, this approach achieves such features without introducing additional symmetries, extra spatial dimensions, or new degrees of freedom beyond those present in the SM.

In our analysis, we focus on fermion species for which the pole-masses can be directly observed—namely, the charged leptons ($e$, $\mu$, $\tau$) and heavy quarks ($c$, $b$, $t$). We do not consider the light quarks ($u$, $d$, $s$), which are subject to significant non-perturbative QCD effects in low-energy observables, as well as the neutrinos, whose masses are sensitive to the neutrino mass mechanism model \cite{Gell-Mann:1979vob, Yanagida:1980xy, PhysRevLett.44.912, PhysRevD.101.115030, Cabrera:2023rcy, CentellesChulia:2024uzv, Mohapatra:1986bd, Gonzalez-Garcia:1988okv, Deppisch:2004fa, PhysRevD.87.113003}.

This paper is organized as follows. In section~\ref{secII}, we introduce the construction of the additive and multiplicative form of Lagrangian. In section~\ref{sec2}, we apply the multiplicative form of the Higgs Lagrangian to electroweak (EW) theory.  In section~\ref{sechiearchymechanism}, we introduce a hierarchical mechanism and, in section~\ref{secfermionhierarchy}, this mechanism is applied to fermion mass. In section~\ref{secunifypoint}, we show that our framework allow the single value of the Yukawa coupling of the charged leptons and heavy quarks. In section~\ref{sec:discussion}, we discuss 1. Pattern of Wilson coefficients,  2. the UV-complete property of this mode in the large field limit. Finally, section~\ref{conclusion} give conclusion.

\section{Additive and multiplicative Lagrangians}\label{secII}

According to the inverse problem of the calculus of variations, the form of a Lagrangian is not uniquely determined; distinct functional forms may lead to the same equations of motion (EOM) for a given physical system. For instance, the EOM for a complex scalar field is expected to reproduce the Klein–Gordon equation,
\begin{align} \label{KG}
\partial_\mu\partial^\mu\phi^\dagger+\frac{\partial V}{\partial\phi}=0,
\end{align}
which is consistent with the relativistic energy–momentum relation,
\begin{align}\label{dispersion}
\omega^2-k^2=m^2.
\end{align}
Here, the potential typically takes the form
$
V = m^2\phi^\dagger\phi + \sum_i \frac{c_i}{n!}(\phi^\dagger\phi)^i$, 
where $c_i$ are coupling constants associated with interaction terms beyond the free-field approximation.

In the simplest case, one may consider an additive form of the Lagrangian,
\begin{align}\label{additiveL}
\mathcal{L} = T(X) + Y(y),
\end{align}
where $ X = \partial_\mu\phi^\dagger\partial^\mu\phi/2 $ and $ y = \phi^\dagger\phi$, with $ T(X)$ and $ Y(y) $ to be determined. Substituting Eq.~\eqref{additiveL} into the Euler–Lagrange equation,
\begin{align}\label{EL0}
\frac{\partial\mathcal{L}}{\partial\phi} - \partial_\mu\left(\frac{\partial\mathcal{L}}{\partial(\partial_\mu\phi)}\right) = 0,
\end{align}
yields
\begin{align}\label{ELa1}
0 = \phi^\dagger \frac{\partial Y}{\partial y}
- \partial_\mu\partial^\mu\phi^\dagger \frac{\partial T}{\partial X}
- \partial_\mu\phi^\dagger\, \partial^\mu X \frac{\partial^2 T}{\partial X^2}.
\end{align}
The third term involves  higher-order derivative interactions and is proportional to $p^4$ in momentum space. In the small-momentum limit, such terms are typically suppressed and can be neglected. The equation then simplifies to
\begin{align}\label{additiveFreeEOM}
0 \simeq \phi^\dagger \frac{\partial Y}{\partial y}
- \partial_\mu\partial^\mu\phi^\dagger \frac{\partial T}{\partial X}.
\end{align}
Substituting Eq.~\eqref{KG} into Eq.~\eqref{additiveFreeEOM}, one obtains
\begin{align}
0 \simeq \phi^\dagger \frac{\partial Y}{\partial y}
+ \frac{\partial V}{\partial\phi} \frac{\partial T}{\partial X}.
\end{align}
Using a separation variable method, one finds
\begin{align}\label{difeqA}
\frac{\partial T}{\partial X} = \phi^\dagger \left(\frac{\partial V}{\partial\phi}\right)^{-1} \frac{\partial Y}{\partial y} = A,
\end{align}
where $A$ is an integration constant. Solving this yields
\begin{align}
T(X) = A X, \qquad Y(y) = -A V(y).
\end{align}
Setting $A=1$, the Lagrangian reduces to the familiar additive form,
\begin{align}\label{KGLs}
\mathcal{L} = X - V = \partial_\mu\phi^\dagger\partial^\mu\phi - V,
\end{align}
which reproduces the Klein–Gordon equation and preserves Lorentz invariance.

Importantly, the Lagrangian need not be restricted to an additive structure. A multiplicative form can be considered,
\begin{align}\label{Lm1}
\mathcal{L} = F(X) f(y),
\end{align}
as suggested in~\cite{surawuttinack2016multiplicative,PhysRevD.106.035020}. Substituting Eq.~\eqref{Lm1} into Eq.~\eqref{EL0} and following similar steps yields the solution
\begin{align}\label{Higgs1}
\mathcal{L}^\pm = \left( \pm \Lambda^4 + \partial_\mu\phi^\dagger\partial^\mu\phi \right) e^{\mp \frac{V}{\epsilon \Lambda^4}},
\end{align}
where \( \Lambda \) is a new mass scale. This expression represents an alternative formulation of the complex scalar field Lagrangian with an exponential structure.

Applications of this class of Lagrangians to particle physics have been explored in the context of the hierarchy problem, neutrino mass generation, and the strong CP problem \cite{PhysRevD.106.035020, Supanyo:2024rzy, Supanyo:2023jkh}. It is important to emphasize, however, that a common resolution of all these issues within a single value of \( \Lambda \) is not achieved. Nonetheless, our recent work \cite{Supanyo:2024rzy} demonstrates that, under certain assumptions, this framework can reduce the apparent hierarchy of Yukawa couplings from \( \mathcal{O}(10^{-6}) \) to \( \mathcal{O}(1) \). In the following section, we apply Eq.~\eqref{Higgs1} to the electroweak sector.

\section{Electroweak model with multiplicative Higgs Lagrangian}\label{sec2}

\begin{figure}[h]
    \centering
    \includegraphics[width=0.5\linewidth]{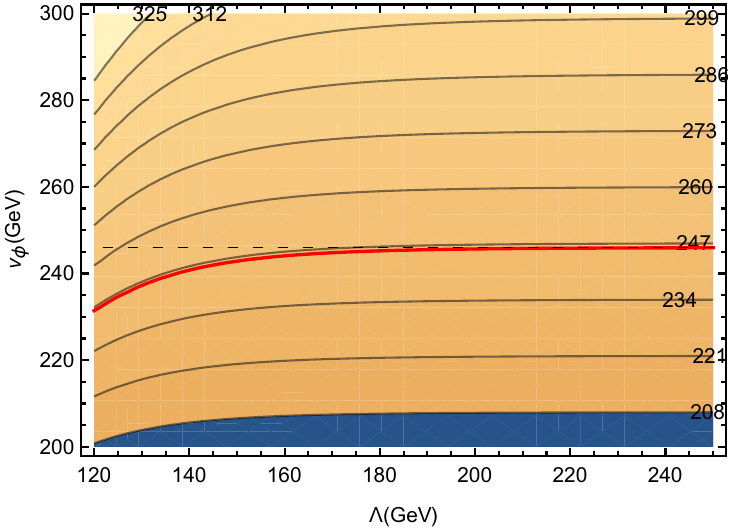}
    \caption{Contour plot of Eq.~\eqref{VEVa} in the $(v_\phi,\Lambda)$ plane. The contour lines represent $v$, with the red line denoting $v=246$ GeV. }
    \label{fig:VEV}
\end{figure}

The EW Lagrangian can be written in the form
\begin{align}
    \mathcal{L}_{EW}=\mathcal{L}_\text{Higgs}+\mathcal{L}_\text{Gauge}+\mathcal{L}_\text{Fermion}.
\end{align}
From Eq.~\eqref{Higgs1}, there are two possible expressions for the Higgs Lagrangian. In this work, we consider their linear combination,
\begin{align}\label{Lsup}
    \mathcal{L}_\text{Higgs}&=\tfrac{1}{2}\mathcal{L}_\text{Higgs}^+ +\tfrac{1}{2}\mathcal{L}_\text{Higgs}^-\nonumber \\
    &=\tfrac{1}{2}\left(+\Lambda^4+(D_\mu\phi)^\dagger D^\mu\phi\right)e^{-V/\Lambda^4}
    +\tfrac{1}{2}\left(-\Lambda^4+(D_\mu\phi)^\dagger D^\mu\phi\right)e^{+V/\Lambda^4},
\end{align}
where the factor $1/2$ is introduced for convenience. In the limit $\Lambda^4\gg V$, the expression reduces to the familiar Higgs sector of the SM,
\begin{align}
    \mathcal{L}_\text{Higgs}\simeq (D_\mu\phi)^\dagger D^\mu\phi - V.
\end{align}
Here $D_\mu$ denotes the covariant derivative associated with SU(2)$\times$U(1), and $V$ is the standard Higgs potential
\begin{align}\label{GL}
    V=-\mu^2\phi^\dagger\phi+\lambda (\phi^\dagger\phi)^2.
\end{align}
From Eq.~\eqref{Lsup}, the effective potential term in the Lagrangian can be identified from the non-derivative part of the Higgs field as
\begin{align}\label{Utree}
    U=\tfrac{\Lambda^4}{2}\left(e^{V/\Lambda^4}-e^{-V/\Lambda^4}\right).
\end{align}
For simplicity, we neglect the Goldstone modes and parametrize
\begin{align}
    \phi=\frac{1}{\sqrt{2}}\begin{pmatrix}
        0\\ \phi_0
    \end{pmatrix}.
\end{align}
Here, in the large field limit, the potential in Eq.~\eqref{Utree} is governed by the positive contribution from the exponential function
\begin{align}
    U\simeq\frac{\Lambda^4}{2}e^{\frac{\lambda \phi_0^4}{4\Lambda^4}},
\end{align}
which ensures the tree-level stability in the large field limit.
On the other hand, in the limit $\phi_b\ll\Lambda$, the potential in Eq.~\eqref{Utree} is reduced in to the standard Higgs potential,
\begin{align}
    U\simeq -\frac{\mu^2\phi_0^2}{2}+\frac{\lambda \phi_0^4}{4}.
\end{align}
The extrema of the full potential in Eq.~\eqref{Utree}  can be evaluated by
\begin{align}
    \frac{\partial U}{\partial\phi_0}=\phi_0 \left(\lambda \phi_0^2-\mu^2\right) 
    \cosh\!\left(\frac{2 \mu^2 \phi_0^2-\lambda \phi_0^4}{4 \Lambda^4}\right)=0.
\end{align}
Since the hyperbolic cosine is always positive for real parameters, the extrema are determined by $\phi_0(\lambda\phi_0^2-\mu^2)=0$. Evaluating the second derivative gives $\partial^2U/\partial\phi_0^2|_{\phi_0=0}=-\mu^2$ and $\partial^2U/\partial\phi_0^2|_{\phi_0=\mu/\sqrt{\lambda}}=2\mu^2\cosh(\mu^4/4\lambda\Lambda^4)$. Thus $\phi_0=0$ corresponds to a local maximum, while
\begin{align}
    |\langle\phi_0\rangle|=\frac{\mu}{\sqrt{\lambda}}\equiv v_\phi
\end{align}
is the minimum of the potential.  Expanding around the vacuum expectation value (VEV),
\begin{align}
    \phi_0=v_\phi+h,
\end{align}
and considering small fluctuations $h\ll \Lambda$, Eq.~\eqref{Lsup} can be written as
\begin{align}\label{Lsup2}
     \mathcal{L}_\text{Higgs}= \mathcal{L}_\text{Higgs}^{(0)}+ \mathcal{L}_\text{Higgs}^{(2)}+ \mathcal{L}_\text{Higgs}^{(3)}+ \mathcal{L}_\text{Higgs}^{(4)}+\dots +\mathcal{L}_\text{gauge-interactions},
\end{align}
where the superscript denotes the order in $h$. The quadratic part is
\begin{align}\label{Lc2}
    \mathcal{L}_\text{Higgs}^{(2)}=\frac{\kappa}{2}\,\partial_\mu h\partial^\mu h- \kappa \mu^2 h^2,
\end{align}
with
\begin{align}\label{beta0}
    \kappa=\tfrac{1}{2}e^{-\mu^2 v_\phi^2/4\Lambda^4}\left(1+e^{\mu^2 v_\phi^2/2\Lambda^4}\right)
    =\cosh\!\left(\frac{\mu^2 v_\phi^2}{4\Lambda^4}\right).
\end{align}
Canonical normalization is restored by redefining
\begin{align}\label{red}
    h\to\kappa^{-1/2} h,
\end{align}
which yields
\begin{align}\label{Lh2}
    \mathcal{L}_\text{Higgs}^{(2)}=\frac{1}{2}\,\partial_\mu h\partial^\mu h-\mu^2 h^2.
\end{align}
The resulting tree-level relations are
\begin{align}\label{massH01}
    \mu^2=\frac{M_h^2}{2},\qquad  \lambda=\frac{M_h^2}{2v_\phi^2}.
\end{align}
The tree-level $W$ and $Z$ masses follow as
\begin{align}
    m_W^2&=\tfrac{1}{4}\kappa g^2 v_\phi^2, \label{mw}\\
    m_Z^2&=\tfrac{1}{4}\kappa v_\phi^2 \left(g^2+g'^2\right).\label{mz}
\end{align}
It is therefore natural to define
\begin{align}\label{VEVa}
    v^2=\kappa v_\phi^2=\cosh\!\left(\frac{M_h^2 v_\phi^2}{8\Lambda^4}\right)v_\phi^2,
\end{align}
where $v=246~\text{GeV}$ \cite{EWtest} denotes the SM Higgs VEV.  Figure~\ref{fig:VEV} shows contour lines of Eq.~\eqref{VEVa} in the $(v_\phi,\Lambda)$ plane. For $\Lambda\gtrsim 180$ GeV, one finds $v_\phi\simeq v$, while for $\Lambda\lesssim 180$ GeV, the framework allows slight deviations of the Higgs VEV with $v_\phi<v$.

In this section we have analyzed the electroweak Higgs sector with a multiplicative Lagrangian. The framework reproduces the SM Higgs structure in the large-$\Lambda$ limit, while at moderate $\Lambda$ it allows small modifications in the Higgs VEV and introduces rescaling factors in the kinetic and mass terms. These effects can be interpreted as deviations from the SM already at the level of dimension-3 and dimension-4 operators. In the next subsections we therefore examine the structure of dimension-3 and dimension-4 operators in comparison to the SM, and subsequently analyze the role of dimension-5 and dimension-6 operators in light of current phenomenological constraints.
\subsection{Constraints on gauge–Higgs couplings, quartic gauge–Higgs couplings, and trilinear Higgs self-couplings}

\begin{figure}[h]
    \centering
    \includegraphics[width=0.8\linewidth]{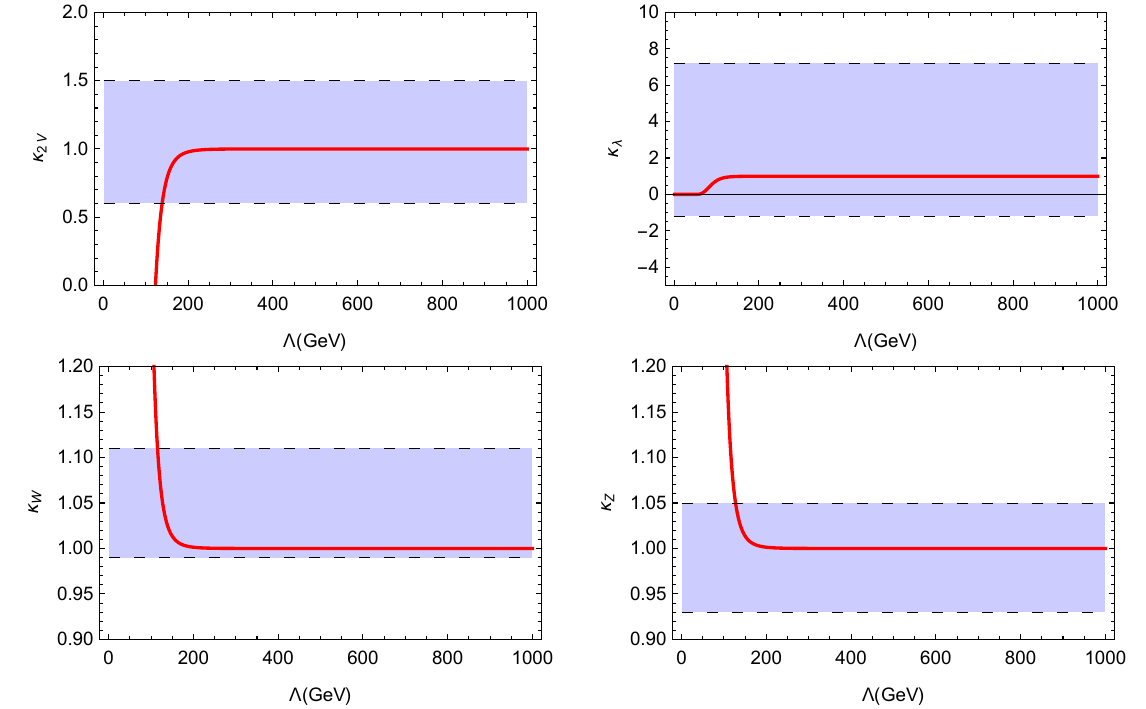}
    \caption{Dependence of the $\kappa_i$ parameters from Eq.~\eqref{kappa} on $\Lambda$, with $M_h=125$ GeV \cite{HiggsMass125,HiggsMass1252} and $v_\phi=246$ GeV\cite{EWtest}. The shaded bands indicate the ATLAS Run-2 constraints: $\kappa_W=1.05^{+0.06}_{-0.06}$, $\kappa_Z=0.99^{+0.06}_{-0.06}$ \cite{Anisha:2022ctm, ATLAS:2022vkf}, and $-1.2<\kappa_\lambda<7.2$, $0.5<\kappa_{2V}<1.5$ \cite{ATLAS:2024fkg}. }
    \label{fig:kappa}
\end{figure}

The parameters $\kappa_i$ quantify the deviations of Higgs and gauge couplings from their SM values. They are introduced through the effective interaction terms
\begin{align}
   -\kappa_\lambda \lambda_{3}^\text{SM} h^3
   -\kappa_W g_{hWW}^\text{SM} h W^\mu_+ W_{\mu}^-
   -\kappa_Z g_{hZZ}^\text{SM} h Z^\mu Z_{\mu}
   -\kappa_{2W} g_{hhWW}^\text{SM} h^2 W^\mu_+ W_{\mu}^-
   -\kappa_{2Z} g_{hhZZ}^\text{SM} h^2 Z^\mu Z_{\mu},
\end{align}
where the rescaling factors follow from the multiplicative Higgs Lagrangian as
\begin{align}\label{kappa}
 &\kappa_\lambda=1/\sqrt{\cosh\left(\frac{M_h^2v_\phi^2}{8\Lambda^4}\right)},~\kappa_W=\kappa_Z=\sqrt{\cosh\left(\frac{M_h^2v_\phi^2}{8\Lambda^4}\right)},\nonumber
 \\
 &\kappa_{2W}=\kappa_{2Z}=1-\frac{M_h^2v_\phi^2}{2\Lambda^4}\tanh\left(\frac{M_h^2v_\phi^2}{8\Lambda^4}\right).
\end{align}
For convenience we denote $\kappa_V=\kappa_W=\kappa_Z$ and $\kappa_{2V}=\kappa_{2W}=\kappa_{2Z}$. The variation of these parameters with $\Lambda$ is shown in figure~\ref{fig:kappa}. For $\Lambda\gtrsim 180$ GeV, corresponding to the regime where $v_\phi\simeq v$, the couplings $\kappa_\lambda$, $\kappa_{2V}$, $\kappa_W$, and $\kappa_Z$ lie within the experimental bounds from ATLAS Run-2 data. By contrast, for $\Lambda\lesssim 180$ GeV, sizable deviations can appear, and the couplings may fall outside the current limits.

The non-additive Higgs Lagrangian modifies the Higgs–gauge and Higgs self-interactions through rescaling factors $\kappa_i$ that depend on $\Lambda$. In the large-$\Lambda$ limit, these reduce to their SM values, while for lower $\Lambda$ significant deviations are possible. The present LHC Run-2 constraints already restrict $\Lambda$ to be above the $\mathcal{O}(100\ \text{GeV})$ scale. These results provide the basis for a more systematic analysis in terms of deviations at the level of dimension-3 and dimension-4 operators, and subsequently through dimension-5 and dimension-6 operators in the SMEFT framework.
\subsection{Dimension-5 and dimension-6 operators}\label{sec3dim}

\begin{figure}[h]
    \centering
    \includegraphics[width=0.5\linewidth]{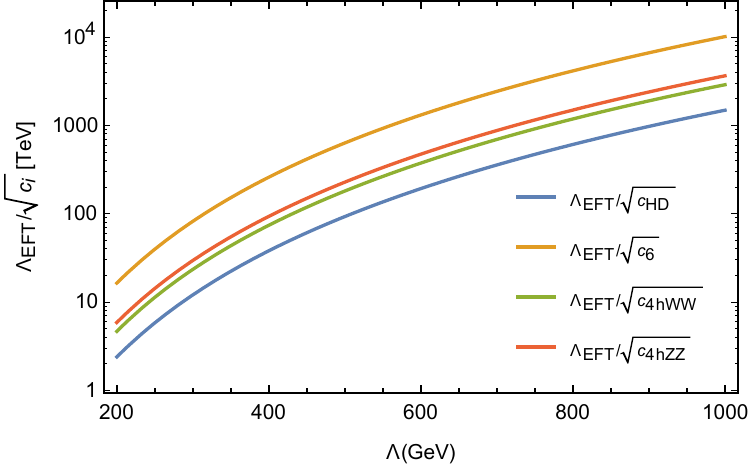}
    \caption{Effective cutoff $\Lambda_\text{EFT}/\sqrt{c_i}$, derived from Eqs.~(36)–(42), as a function of $\Lambda$.}
    \label{fig:eftfig}
\end{figure}

From Eq.~\eqref{Lsup}, the effective Lagrangian in the broken phase generates higher-dimensional operators of the form
\begin{align}
   \mathcal{L}_{\dim-5} &=\frac{c_5}{\Lambda_\text{EFT}}h^5
   +\frac{c_{3hWW}}{\Lambda_\text{EFT}}h^3W^\mu_+W_\mu^-
   +\frac{c_{3hZZ}}{\Lambda_\text{EFT}}h^3Z_\mu Z^\mu, \label{L5}\\
   \mathcal{L}_{\dim-6} &=\frac{c_{HD}}{\Lambda_\text{EFT}^2}h^2\partial_\mu h\partial^\mu h
   +\frac{c_6}{\Lambda_\text{EFT}^2}h^6
   +\frac{c_{4hWW}}{\Lambda_\text{EFT}^2}h^4W^\mu_+W_\mu^-
   +\frac{c_{4hZZ}}{\Lambda_\text{EFT}^2}h^4Z^\mu Z_\mu, \label{L6}
\end{align}
where $\Lambda_\text{EFT}$ is the EFT cutoff scale and $c_i$ denote the corresponding Wilson coefficients defined in Eqs.~\eqref{c5}-\eqref{ch4zz}. For simplicity, we apply $v_\phi\simeq v$ as
\begin{align}
    &c_5=\frac{\Lambda_\text{EFT} M_h^4 \sinh\left(\frac{M_h^2v^2}{8\Lambda^4}\right)}{4v\Lambda^4\cosh\left(\frac{M_h^2v^2}{8\Lambda^4}\right)^{5/2}},\label{c5}
    \\
    &c_{3hWW}=-\frac{3\Lambda_\text{EFT}g^2M_h^2v\sinh\left(\frac{M_h^2v^2}{8\Lambda^4}\right)}{8\Lambda^4\cosh\left(\frac{M_h^2v^2}{8\Lambda^4}\right)^{3/2}},
    \\
    &c_{3hZZ}=-\frac{3\Lambda_\text{EFT}(g^2+g'^2)M_h^2v\sinh\left(\frac{M_h^2v^2}{8\Lambda^4}\right)}{16\Lambda^4\cosh\left(\frac{M_h^2v^2}{8\Lambda^4}\right)^{3/2}},
    \\
    &c_{HD}=-\frac{\Lambda_\text{EFT}^2M_h^2\tanh\left(\frac{M_h^2v^2}{8\Lambda^4}\right)}{4\Lambda^4\cosh\left(\frac{M_h^2v^2}{8\Lambda^4}\right)},
    \\
   & c_6=-\frac{\Lambda_\text{EFT}^2M_h^4(M_h^2v^2-9\Lambda^4\tanh\left(\frac{M_h^2v^2}{8\Lambda^4}\right))}{48v^2\Lambda^8\cosh\left(\frac{M_h^2v^2}{8\Lambda^4}\right)^{2}},
   \\  &c_{4hWW}=\frac{g^2\Lambda_\text{EFT}^2M_h^2(M_h^2v^2-13\Lambda^4\tanh\left(\frac{M_h^2v^2}{8\Lambda^4}\right))}{32\Lambda^8\cosh\left(\frac{M_h^2v^2}{8\Lambda^4}\right)},
   \\  &c_{4hZZ}=\frac{(g^2+g'^2)\Lambda_\text{EFT}^2M_h^2(M_h^2v^2-13\Lambda^4\tanh\left(\frac{M_h^2v^2}{8\Lambda^4}\right))}{64\Lambda^8\cosh\left(\frac{M_h^2v^2}{8\Lambda^4}\right)},\label{ch4zz}
\end{align}
From global SMEFT fits to Higgs, diboson, and electroweak precision data constrain the coefficient $\bar{C}_{HD}$ in the operator
\begin{align}\label{HDH}
    \frac{\bar{C}_{HD}}{v^2}|H^\dagger D_\mu H|^2,
\end{align}
with $|\bar{C}_{HD}|\lesssim 0.001$ \cite{Ellis:2018gqa}, where, in this reference, $\Lambda_\text{EFT}=v$. In the broken phase this operator contributes the SM as
\begin{align}\label{warsaw}
    \mathcal{L}_{\dim-6}\supset\frac{\bar{C}_{HD}}{2v^2}h^2\partial_\mu h\partial^\mu h.
\end{align}
By tree-level matching between Eq.~\eqref{L6} and Eq.~\eqref{warsaw}, we obtain
\begin{align}
    \bar{C}_{HD} =2c_{HD}\frac{v^2}{\Lambda_\text{EFT}^2}
    =\frac{M_h^2v^2\tanh\!\left(\tfrac{M_h^2v^2}{8\Lambda^4}\right)}
    {2\Lambda^4\cosh\!\left(\tfrac{M_h^2v^2}{8\Lambda^4}\right)}.
\end{align}
Imposing the phenomenological bound $|\bar{C}_{HD}|<0.001$ requires
\begin{align}\label{SMEFTconstraint}
    \Lambda \gtrsim 294~\text{GeV}.
\end{align}
In this parameter range, the cutoff scale of the multiplicative Higgs model lies above $\mathcal{O}(10)$~TeV, as illustrated in figure~\ref{fig:eftfig}.

In this subsection, we have derived the dimension-5 and dimension-6 operators induced by the multiplicative Higgs Lagrangian. The resulting Wilson coefficients are expressed explicitly in terms of $\Lambda$, and tree-level matching to SMEFT allows comparison with global constraints. In particular, the stringent bound on $\bar{C}_{HD}$ implies $\Lambda\gtrsim 294$ GeV, corresponding to an effective cutoff above $\mathcal{O}(10)$ TeV. These constraints indicate that the multiplicative Higgs model remains consistent with current data only in a restricted parameter space $\Lambda\gtrsim 294~\text{GeV}$.

In the next section we will explore whether this framework can provide a viable mechanism to address the fermion mass hierarchy.

\section{The hierarchy mechanism}\label{sechiearchymechanism}

In the standard gauge theory, the EW Lagrangian is constructed as a linear combination of gauge-invariant operators $O_g$,  
\begin{align}\label{sumggg}
    \mathcal{L}_\text{EW}=\sum_i O_{g,i},
\end{align}
where $O_g$ includes Yukawa interactions as well as the covariant kinetic terms for gauge, scalar, and fermion fields. In the multiplicative Higgs framework, however, gauge-invariant operators need not enter only as a simple linear sum. Instead, they can appear in exponential structures while preserving gauge invariance, leading to  
\begin{align}\label{Lsupgaugex}
    \mathcal{L}=& \epsilon_1 O_g+\tfrac{1}{2}\Big(+\Lambda^4+(D_\mu\phi)^\dagger D^\mu\phi+\epsilon_2 O_g\Big)e^{-\frac{V+\epsilon_3 O_g}{\Lambda^4}} +\tfrac{1}{2}\Big(-\Lambda^4+(D_\mu\phi)^\dagger D^\mu\phi+\epsilon_4 O_g\Big)e^{+\frac{V+\epsilon_5 O_g}{\Lambda^4}},
\end{align}
where $\epsilon_i=\{0,\pm1\}$ specify the inclusion pattern of $O_g$.  After SSB, the Higgs field is expanded around its VEV as  
\begin{align}
    \phi^\dagger = \frac{1}{\sqrt{2}}\,(0~~v+\kappa^{-1/2}h).
\end{align}
Expanding the Lagrangian in the limit $\Lambda^4\gg O_g$ and $8\Lambda^4\gg M_h^2v^2$, the contribution containing only $O_g$ becomes  
\begin{align}
    \mathcal{L}_{O_g}= f\,O_g+\mathcal{O}(O_g^2),
\end{align}
where the scaling factor $f$ depends on the choice of $\epsilon=(\epsilon_1,\epsilon_2,\epsilon_3,\epsilon_4,\epsilon_5)$ as  
\begin{align}\label{f000}
  f\simeq \tfrac{1}{2}\Big(2\epsilon_1+\epsilon_2-\epsilon_3+\epsilon_4-\epsilon_5\Big) 
   +\frac{(-\epsilon_2+\epsilon_3+\epsilon_4-\epsilon_5)M_h^2v^2}{16 \Lambda^4}
   +\frac{(\epsilon_2-\epsilon_3+\epsilon_4-\epsilon_5)M_h^4v^4}{256 \Lambda^8}.
\end{align}
Formally, there are $3^5=243$ possible configurations of $\epsilon_i$. After eliminating redundancies, nine distinct leading-order scaling factors remain,  
\begin{align}\label{fa}
    &|f_{(-8)}|=\frac{M_h^4 v^4}{128\Lambda^8},~~|f_{(-4)}^{(1)}|=\frac{M_h^2v^2}{8\Lambda^4},~~|f_{(-4)}^{(2)}|=\frac{M_h^2v^2}{4\Lambda^4}, \nonumber \\
    &|f_{(0)}^{(1)}|=\tfrac{1}{2},~~|f_{(0)}^{(2)}|=1,~~|f_{(0)}^{(3)}|=\tfrac{3}{2},~~|f_{(0)}^{(4)}|=2,~~|f_{(0)}^{(5)}|=\tfrac{5}{2},~~|f_{(0)}^{(6)}|=3.
\end{align}
We denote these factors as  
\begin{align}\label{ffactordef}
    f= f_{(\text{order in }\Lambda)}^{(i)},
\end{align}
with $(i)$ labeling degenerate values at a given order of $\Lambda$.  Importantly, negative values of $f_{(n)}$ are excluded. For instance, if $O_g$ represents the fermion kinetic term, a negative coefficient would signal ghost-like instabilities and hence is excluded. Including next-to-leading terms in the expansion further splits the scaling factors in Eq.~\eqref{fa} into sub-branches as follows.
\begin{align}
    &f_{(-8)}=\frac{M_h^4v^4}{128\Lambda^8}\left(1+\frac{M_h^4v^4}{768\Lambda^8}\right)\label{fm8}
    \\
    &f_{(-4)}^{(1)}\to\begin{cases}
        f_{(-4)}^{(1,1)}=\frac{M_h^2v^2}{8\Lambda^4}\left(1-\frac{M_h^2v^2}{16\Lambda^4}\right)
        \\
        f_{(-4)}^{(1,2)}=\frac{M_h^2v^2}{8\Lambda^4}\left(1+\frac{M_h^4v^4}{128\Lambda^8}\right)
        \\
        f_{(-4)}^{(1,3)}=\frac{M_h^2v^2}{8\Lambda^4}\left(1+\frac{M_h^2v^2}{16\Lambda^4}\right)
    \end{cases}
    \\
    &f_{(-4)}^{(2)}=\frac{M_h^2v^2}{4\Lambda^4}\left(1+\frac{M_h^4v^4}{384\Lambda^8}\right)
\end{align}
\begin{align}
   & f_{(0)}^{(1)}=\frac{1}{2}\to\begin{cases}
        f_{(0)}^{(1,1)}=\frac{1}{2}\left(1-\frac{3M_h^2v^2}{8\Lambda^4}\right)
        \\
        f_{(0)}^{(1,2)}=\frac{1}{2}\left(1-\frac{M_h^2v^2}{8\Lambda^4}\right)
        \\
         f_{(0)}^{(1,3)}=\frac{1}{2}\left(1+\frac{M_h^2v^2}{8\Lambda^4}\right)
        \\
        f_{(0)}^{(1,4)}=\frac{1}{2}\left(1+\frac{3M_h^2v^2}{8\Lambda^4}\right)
    \end{cases}
\end{align}
\begin{align}
    & f_{(0)}^{(2)}=1\to \begin{cases}
        f_{(0)}^{(2,1)}=1-\frac{M_h^2v^2}{4\Lambda^4}
        \\
        f_{(0)}^{(2,2)}=1-\frac{M_h^2v^2}{8\Lambda^4}
        \\
        f_{(0)}^{(2,3)}=1
        \\
        f_{(0)}^{(2,4)}=1+\frac{M_h^4v^4}{128\Lambda^8}
        \\
        f_{(0)}^{(2,5)}=1+\frac{M_h^4v^4}{64\Lambda^8}
        \\
        f_{(0)}^{(2,6)}=1+\frac{M_h^2v^2}{8\Lambda^4}
        \\
        f_{(0)}^{(2,7)}=1+\frac{M_h^2v^2}{4\Lambda^4}
    \end{cases}
\end{align}
\begin{align}
    f_{(0)}^{(3)}=\frac{3}{2}\to\begin{cases}
        f_{(0)}^{(3,1)}=\frac{3}{2}\left(1-\frac{M_h^2v^2}{8\Lambda^4}\right)
        \\
        f_{(0)}^{(3,2)}=\frac{3}{2}\left(1-\frac{M_h^2v^2}{24\Lambda^4}\right)
        \\
         f_{(0)}^{(3,3)}=\frac{3}{2}\left(1+\frac{M_h^2v^2}{24\Lambda^4}\right)
        \\
        f_{(0)}^{(3,4)}=\frac{3}{2}\left(1+\frac{M_h^2v^2}{8\Lambda^4}\right)
    \end{cases}
\end{align}
\begin{align}
    f_{(0)}^{(4)}=2\to\begin{cases}
        f_{(0)}^{(4,1)}=2\left(1-\frac{M_h^2v^2}{16\Lambda^4}\right)
        \\
        f_{(0)}^{(4,2)}=2\left(1+\frac{M_h^4v^4}{256\Lambda^8}\right)
        \\
        f_{(0)}^{(4,3)}=2\left(1+\frac{M_h^4v^4}{128\Lambda^8}\right)
        \\
         f_{(0)}^{(4,4)}=2\left(1+\frac{M_h^2v^2}{16\Lambda^4}\right)
    \end{cases}
\end{align}
\begin{align}
  &  f_{(0)}^{(5)}=\frac{5}{2}\to\begin{cases}
        f_{(0)}^{(5,1)}=\frac{5}{2}\left(1-\frac{M_h^2v^2}{40\Lambda^4}\right)
        \\
        f_{(0)}^{(5,2)}=\frac{5}{2}\left(1+\frac{M_h^2v^2}{40\Lambda^4}\right)
    \end{cases}
    \\
    &f_{(0)}^{(6)}=3\left(1+\frac{M_h^4 v^4}{192\Lambda^8}\right)\label{f06}
\end{align}
We have introduced a new index $f_{(n)}^{(i)}\to f_{(n)}^{(i,j)}$ to specify the sub-scale for $f_{(n)}^{(i)}$. Here, the associated configurations of $(\epsilon_1,\epsilon_2,\epsilon_3,\epsilon_4,\epsilon_5)$ to provide the corresponding  $f_{(n)}^{(i,j)}$ are listed in appendix \ref{AppA}. 

\medskip
 In this section, we have shown that gauge-invariant operators $O_g$ can acquire nontrivial scaling factors $f_{(n)}^{(i,j)}$ when embedded in the multiplicative Higgs framework. These scaling factors range from suppressed values $\propto 1/\Lambda^n$ to $\mathcal{O}(1)$, while negative values are excluded due to stability considerations. The classification of possible $f_{(n)}^{(i,j)}$ provides a systematic mechanism to generate hierarchical effective couplings. In the next step, we will incorporate this mechanism into the fermion sector to investigate whether it can account for the observed fermion mass hierarchy.
\section{Fermion Hierarchy}\label{secfermionhierarchy}

The left-handed fermion doublet and the right-handed fermion singlet are introduced as
\begin{align}\label{fermionfield}
    \Psi_L=\begin{pmatrix}
        \chi_L
        \\
        \psi_L
    \end{pmatrix},~\chi_R~\text{and}~\psi_R,
\end{align}
where $\Psi_L$ transforms as an $SU(2)_L$ doublet, while $\chi_R$ and $\psi_R$ are singlets.  The gauge-invariant operators $O_g$ associated with these fields are
\begin{align}
    &O^{L}=\overline{\Psi}_L i\gamma^\mu D_\mu\Psi_L,~
    O^{R,\psi}=\overline{\psi}_R i\gamma^\mu D_\mu \psi_R,~
    O^{R,\chi}=\overline{\chi}_R i\gamma^\mu D_\mu \chi_R,\nonumber \\
    &O^{Y,\psi}=-\lambda_\psi \overline{\Psi}_L\phi\psi_R+h.c.,~
    O^{Y,\chi}=-\lambda_\chi\overline{\Psi}_L\tilde{\phi}\chi_R+h.c.,
\end{align}
where $\tilde{\phi}^\dagger=(\phi_0~0)/\sqrt{2}$. In the SM context, $\chi$ corresponds to $\nu_e$, $\nu_\mu$, $\nu_\tau$, $u$, $c$, and $t$, while $\psi$ corresponds to $e$, $\mu$, $\tau$, $d$, $s$, and $b$.  The fermion Lagrangian coupled to the Higgs sector can be written in the compact form
\begin{align}\label{Lsupgauge2}
    \mathcal{L}= O^\Psi_1+\frac{1}{2}(+\Lambda^4+(D_\mu\phi)^\dagger D^\mu\phi+ O^\Psi_2)e^{-\frac{V+ O^\Psi_3}{\Lambda^4}}
    +\frac{1}{2}(-\Lambda^4+(D_\mu\phi)^\dagger D^\mu\phi+ O^\Psi_4)e^{+\frac{V+ O^\Psi_5}{\Lambda^4}},
\end{align}
with
\begin{align}
   O^\psi_i=& \epsilon^L_i O_L+\epsilon^{R,\psi}_i O_{R,\psi}+\epsilon^{R,\chi}_i O_{R,\chi}+\epsilon^{Y,\chi}_i O_{Y,\chi}+\epsilon^{Y,\psi}_i O_{Y,\psi}.
\end{align}
Here, the subscript of $\epsilon_i^{j}$ specifies the position of $O_j$, while the superscript refers to the operator type. After SSB, the quadratic part of the fermion Lagrangian becomes
\begin{align}\label{psi0}
    \mathcal{L}_{\psi/\chi}^{(2)}=&f_{L,(n_L)}^{(i_L)}\overline{\psi}_L i\gamma^\mu\partial_\mu \psi_L
    +f_{R,(n_R)}^{(i_R)}\overline{\psi}_R i\gamma^\mu\partial_\mu \psi_R
    -f_{Y,(n_Y)}^{(i_Y)}\frac{\lambda_\psi v_\phi}{\sqrt{2}}(\overline{\psi}_L\psi_R +h.c.)\nonumber \\
    &+f_{L,(n_L)}^{(i_L)}\overline{\chi}_L i\gamma^\mu\partial_\mu \chi_L
    +f_{R,(m_R)}^{(j_R)}\overline{\chi}_R i\gamma^\mu\partial_\mu \chi_R
    -f_{Y,(m_Y)}^{(k_Y)}\frac{\lambda_\chi v_\phi}{\sqrt{2}}(\overline{\chi}_L\chi_R +h.c.).
\end{align}
Canonical normalization kinetic energy can be obtained through the field redefinitions
\begin{align}\label{redpsi}
    \psi_L\to\frac{\psi_L}{\sqrt{f_{L,(n_L)}^{(i_L)}}},~
    \psi_R\to\frac{\psi_R}{\sqrt{f_{R,(n_R)}^{(i_R)}}},~
    \chi_L\to\frac{\chi_L}{\sqrt{f_{L,(n_L)}^{(i_L)}}},~
    \chi_R\to\frac{\chi_R}{\sqrt{f_{R,(m_R)}^{(j_R)}}},
\end{align}
leading to
\begin{align}\label{psi1}
    \mathcal{L}_\psi^{(2)}=&\overline{\psi}_L i\gamma^\mu\partial_\mu \psi_L
    +\overline{\psi}_R i\gamma^\mu\partial_\mu \psi_R
    -\frac{f_{Y,(n_Y)}^{(i_Y)}}{\sqrt{f_{L,(n_L)}^{(i_L)}f_{R,(n_R)}^{(i_R)}}}\frac{\lambda_\psi v_\phi}{\sqrt{2}}(\overline{\psi}_L\psi_R +h.c.)\nonumber \\
    &+\overline{\chi}_L i\gamma^\mu\partial_\mu \chi_L
    +\overline{\chi}_R i\gamma^\mu\partial_\mu \chi_R
    -\frac{f_{Y,(m_Y)}^{(k_Y)}}{\sqrt{f_{L,(n_L)}^{(i_L)}f_{R,(m_R)}^{(j_R)}}}\frac{\lambda_\chi v_\phi}{\sqrt{2}}(\overline{\chi}_L\chi_R +h.c.).
\end{align}
The tree-level fermion masses are then expressed as
\begin{align}\label{massspectrum1}
    m^\psi_{\left(n_y-\frac{n_L}{2}-\frac{n_R}{2}\right)}
    &=\frac{f_{Y,(n_Y)}^{(i_Y)}}{\sqrt{f_{L,(n_L)}^{(i_L)}f_{R,(n_R)}^{(i_R)}}}\frac{\lambda_\psi v_\phi}{\sqrt{2}}, \\
     m^\chi_{\left(m_y-\frac{n_L}{2}-\frac{m_R}{2}\right)}
     &=\frac{f_{Y,(m_Y)}^{(k_Y)}}{\sqrt{f_{L,(n_L)}^{(i_L)}f_{R,(m_R)}^{(j_R)}}}\frac{\lambda_\chi v_\phi}{\sqrt{2}}.
\end{align}
Due to the $SU(2)_L$ structure, both $m^\psi$ and $m^\chi$ share the same factor $f_L$, while $f_R^\psi$, $f_R^\chi$, $f_Y^\psi$, and $f_Y^\chi$ can vary independently. From Eq.~\eqref{massspectrum1}, the fermion masses can be organized into nine distinct scales as follows.  
\begin{align}
    &m_{(-8)}=m_{(-8-\frac{0}{2}-\frac{0}{2})}~~~~~~\sim \frac{M_h^4v^4}{\Lambda^8}\frac{\lambda_y v}{\sqrt{2}}\label{mm8}
    \\
   & m_{(-6)}=\begin{cases}
       m_{(-8+\frac{4}{2}-\frac{0}{2})}
       \\
       m_{(-8-\frac{0}{2}-\frac{4}{2})}
   \end{cases} \sim\frac{M_h^3 v^3}{\Lambda^6}\frac{\lambda_y v}{\sqrt{2}}
   \\
   & m_{(-4)}=\begin{cases}
       m_{(-8+\frac{4}{2}+\frac{4}{2})}
       \\
        m_{(-4+\frac{0}{2}+\frac{0}{2})}
   \end{cases}
   \sim \frac{M_h^2 v^2}{\Lambda^4}\frac{\lambda_y v}{\sqrt{2}}
   \\
   & m_{(-2)}=\begin{cases}
    m_{(-8+\frac{4}{2}+\frac{8}{2})}
    \\
    m_{(-8+\frac{8}{2}+\frac{4}{2})}
    \\
        m_{(-4+\frac{4}{2}+\frac{0}{2})}
        \\
        m_{(-4+\frac{0}{2}+\frac{4}{2})}
   \end{cases}\sim \frac{M_h v}{\Lambda^2}\frac{\lambda_y v}{\sqrt{2}}
   \\
   & m_{(0)}=\begin{cases}
    m_{(-8+\frac{8}{2}+\frac{8}{2})}
    \\
        m_{(-4+\frac{4}{2}+\frac{4}{2})}
        \\
        m_{(-4+\frac{0}{2}+\frac{0}{2})}
   \end{cases}~~\sim \frac{\lambda_y v}{\sqrt{2}}
  \\
   & m_{(+2)}=\begin{cases}
    m_{(-4+\frac{4}{2}+\frac{8}{2})}
    \\
     m_{(-4+\frac{8}{2}+\frac{4}{2})}
    \\
        m_{(0+\frac{4}{2}+\frac{0}{2})}
        \\
        m_{(0+\frac{0}{2}+\frac{4}{2})}
   \end{cases}\sim \frac{\Lambda^2}{M_h v}\frac{\lambda_y v}{\sqrt{2}}
   \\
  & m_{(+4)}=\begin{cases}
    m_{(-4+\frac{8}{2}+\frac{8}{2})}
    \\
        m_{(0+\frac{4}{2}+\frac{4}{2})}
         \\
        m_{(0+\frac{0}{2}+\frac{8}{2})}
        \\
        m_{(0+\frac{8}{2}+\frac{0}{2})}
   \end{cases}\sim \frac{\Lambda^4}{M_h^2 v^2}\frac{\lambda_y v}{\sqrt{2}}
   \\
   & m_{(+6)}=\begin{cases}
        m_{(0+\frac{4}{2}+\frac{8}{2})}
         \\
        m_{(0+\frac{8}{2}+\frac{4}{2})}
   \end{cases}~\sim \frac{\Lambda^6}{M_h^3 v^3}\frac{\lambda_y v}{\sqrt{2}}
   \\
   & m_{(+8)}=
        m_{(0+\frac{8}{2}+\frac{8}{2})}
 ~~~~~~~\sim \frac{\Lambda^8}{M_h^4 v^4}\frac{\lambda_y v}{\sqrt{2}}\label{mp8}
\end{align}
At the special point $\Lambda=\sqrt{M_h v}\simeq 175~\text{GeV}$, however, all fermion mass scales collapse to $m\sim \lambda_y v/\sqrt{2}$, eliminating the intended hierarchy. This indicates that the mechanism, by itself, does not naturally generate fermion mass hierarchies unless the condition
\begin{align}
    \Lambda>\sqrt{M_h v},
\end{align}
is satisfied. This requirement ensures the separation into nine distinct mass scales while remaining consistent with SMEFT bounds in Eq.~\eqref{SMEFTconstraint}. Thus, the framework still retains the capacity to accommodate fermion mass hierarchies.  

\subsection{Deviation of the Tree-Level Yukawa Interaction}

The Higgs Lagrangian and Yukawa sector in the broken phase read
\begin{align}\label{Ly0}
     \mathcal{L}=\frac{\kappa}{2}\partial_\mu h\partial^\mu h-\kappa\mu^2h^2
     -\frac{f_{Y,(n_Y)}^{(i_Y)}}{\sqrt{f_{L,(n_L)}^{(i_L)}f_{R,(n_R)}^{(i_R)}}}\frac{\lambda_\psi (v_\phi+h)}{\sqrt{2}}\overline{\psi}_L\psi_R 
     -\frac{f_{Y,(m_Y)}^{(k_Y)}}{\sqrt{f_{L,(n_L)}^{(i_L)}f_{R,(m_R)}^{(j_R)}}}\frac{\lambda_\chi (v_\phi+h)}{\sqrt{2}}\overline{\chi}_L\chi_R,
\end{align}
where the fermion redefinitions in Eq.~\eqref{redpsi} are applied. After canonical normalization of the Higgs field, one obtains
\begin{align}
      \mathcal{L}=&\frac{1}{2}\partial_\mu h\partial^\mu h-\mu^2h^2
      -m^{\psi}_{(n_Y-\frac{n_L}{2}-\frac{n_R}{2})}\overline{\psi}_L\psi_R 
      -\frac{m^{\psi}_{(n_Y-\frac{n_L}{2}-\frac{n_R}{2})}}{v}h\overline{\psi}_L\psi_R \nonumber\\
      &-m^{\chi}_{(m_Y-\frac{n_L}{2}-\frac{m_R}{2})}\overline{\chi}_L\chi_R
      -\frac{m^{\chi}_{(m_Y-\frac{n_L}{2}-\frac{m_R}{2})}}{v} h\overline{\chi}_L\chi_R.
\end{align}
Thus, the Yukawa couplings governing $h\to \overline{\psi}\psi$ and $h\to \overline{\chi}\chi$ interactions are
\begin{align}
     y_\psi=\frac{m^{\psi}_{(n_Y-\frac{n_L}{2}-\frac{n_R}{2})}}{v},\quad
     y_\chi=\frac{m^{\chi}_{(m_Y-\frac{n_L}{2}-\frac{m_R}{2})}}{v}.
\end{align}
At tree level, these coincide with the SM values, $\kappa_Y=y/y_\text{SM}=1$.

In summary, this section has shown that embedding fermion kinetic and Yukawa operators in a nonstandard Lagrangian naturally organizes masses into nine hierarchical scales, provided the condition $\Lambda>\sqrt{M_h v}$ is satisfied. This requirement prevents the collapse of all fermion masses into a single scale and remains consistent with SMEFT constraints, leaving viable parameter space for addressing the observed mass spectrum. Importantly, the framework preserves the tree-level Standard Model relation $y_f=m_f/v$, implying no deviation in Yukawa couplings at leading order. In the following section, we demonstrate that this mechanism can reduces Yukawa couplings to $\mathcal{O}(1)$ and also opens the possibility to align the coupling $\lambda$ of the charged leptons and heavy quarks into a single value.

\section{The heavy quarks and charged leptons}\label{secunifypoint}
In this section, we demonstrate that the proposed mechanism  allows Yukawa couplings to be expressed at the $\mathcal{O}(1)$ scale and possibly provides a framework in which the Yukawa couplings of charged leptons and heavy quarks can be equal. In this section, we are base on the on-shell subtraction scheme where the fermion masses in the Lagrangian are observed pole masses.

\begin{figure}[h!]
\centering
\includegraphics[width=1\linewidth]{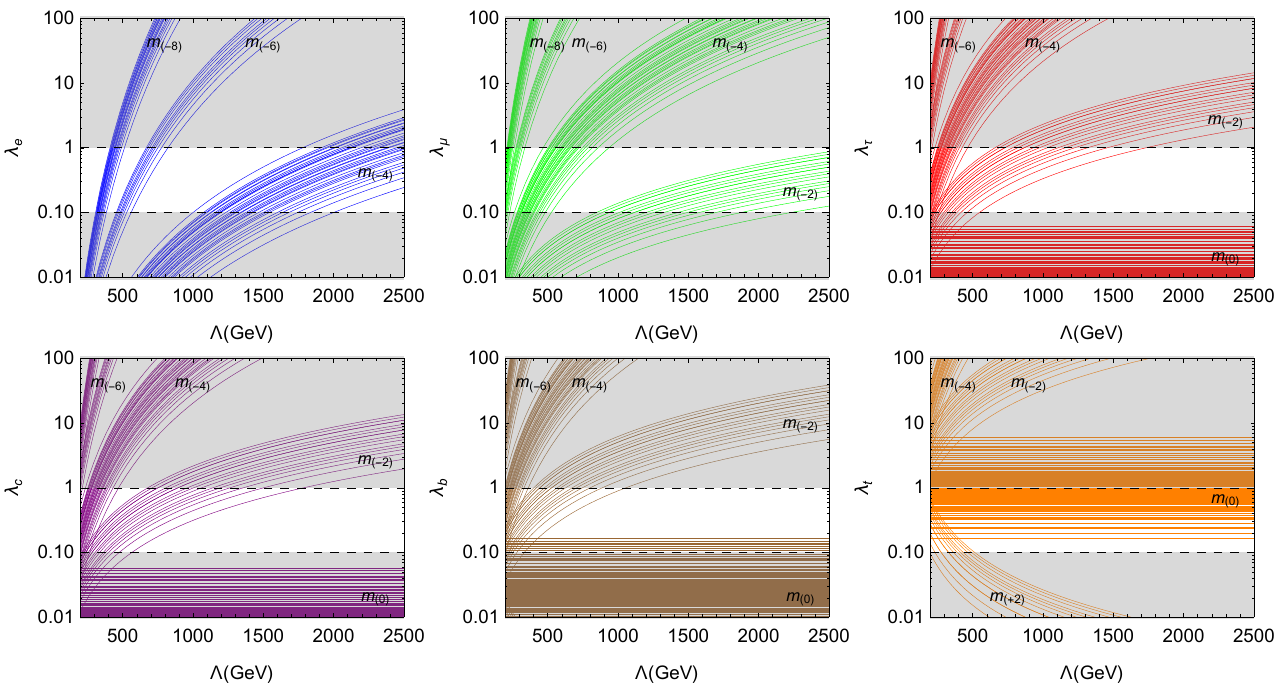}
\caption{Yukawa couplings obtained from Eq.\eqref{Ye} as functions of $\Lambda$, where the central values of Eq.\eqref{leptonmass} are used together with the leading-order scaling factor of Eq.~\eqref{fa}.}
\label{fig:1}
\end{figure}

We adopt the pole masses of the charged leptons and heavy quarks from the Particle Data Group (PDG) \cite{ParticleDataGroup:2024cfk}:
\begin{align}\label{leptonmass}
&M_e=0.5109989500(15)\text{MeV}, \quad M_\mu=105.6583755(23)\text{MeV}, \quad M_\tau=1776.93(09)\text{MeV}, \nonumber 
\\
&M_c=1.67(07)\text{GeV}, \quad M_b=4.78(06)\text{GeV}, \quad M_t=172.56(31)\text{GeV}.
\end{align}
Substituting Eq.\eqref{leptonmass} into Eq.\eqref{massspectrum1}, the original parameters of Yukawa couplings in the model are expressed as
\begin{align}\label{Ye}
    \lambda_{\alpha,(-n_Y+\frac{n_L}{2}+\frac{n_R}{2})}=\frac{\sqrt{2f^{(i_L)}_{L,(n_L)}f^{(i_R)}_{R,(n_R)}}}{f_{Y,(n_Y)}^{(i_Y)}}\frac{M_{\alpha,(n_Y-\frac{n_L}{2}-\frac{n_R}{2})}}{v}
\end{align}
with $\alpha = e, \mu, \tau, c, b, t$. Taking $v\simeq 246$ GeV and $M_h\simeq 125$ GeV, the resulting Yukawa couplings as functions of $\Lambda$ are shown in figure~\ref{fig:1}. The parameter region $\Lambda \lesssim 300\text{GeV}$ is not consider since it conflicts with the lower bound $\Lambda > 294~\text{GeV}$ from phenomenological constraints.
Within this framework, the observed fermion masses can be systematically accommodated by associating them with specific mass levels generated by the mechanism. As a consequence, the corresponding Yukawa couplings remain confined within the natural $\mathcal{O}(1)$ range, where 
\begin{align}
    &M_e\subset\{m_{(-8)},m_{(-6)},m_{(-4)}\},\label{cone}
    \\
    &M_\mu\subset\{m_{(-6)},m_{(-4)}\},
    \\
    &M_\tau\subset\{m_{(-4)},m_{(-2)}\},
    \\
    &M_c\subset\{m_{(-4)},m_{(-2)}\},
    \\
    &M_b\subset\{m_{(-4)},m_{(-2)},m_{(0)}\},
    \\
    &M_t\subset\{m_{(0)},m_{(+2)}\}.\label{cont}
\end{align}
If one is interested in scenarios where the couplings are arranged to be of order $\mathcal{O}(1)$, the model allows for constructing the Lagrangian of these six particle species, as detailed in appendix A and Eqs.~\eqref{mm8}-\eqref{mp8}, in as many as $2\times 10^{32}$ distinct configurations \footnote{Astonishingly, this value happens to be numerically close to the ratio $(M_{\mathrm{Planck}}/(v/\sqrt{2}))^2=(2.44\times10^{18}/174)^2 \simeq 2\times10^{32}$.}. In our opinion, this enormous multiplicity indicates that the $\mathcal{O}(1)$ assumption may not be as natural as commonly expected. In what follows, we aim to narrow down these possibilities by posing the question of whether, within our framework, it is possible to reproduce the observed masses of the charged leptons and heavy quarks while assuming identical original Yukawa coupling parameters.

\begin{figure}[h!]
    \centering
    \includegraphics[width=0.5\linewidth]{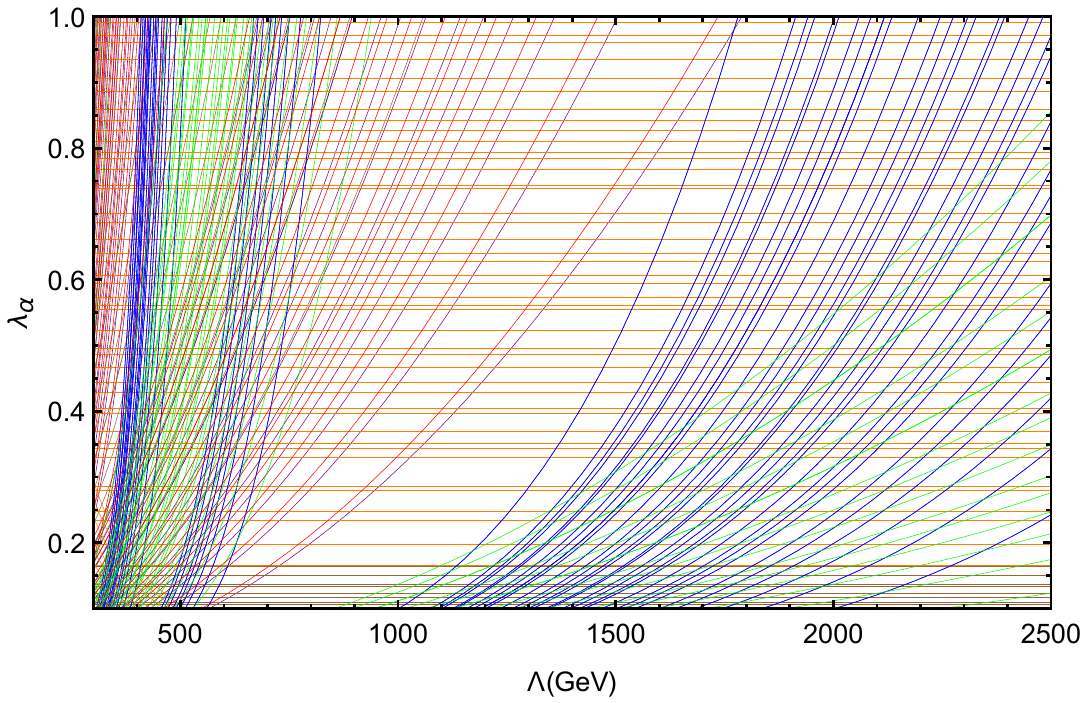}
    \caption{The plot of Eq.~\eqref{Ye} versus $\Lambda$, where $\alpha=e,\mu,\tau,c,b,t$. We use the central values of Eq.~\eqref{leptonmass}, the leading-order scaling factors from Eq.~\eqref{fa}, $M_h\simeq125$ GeV, and $v\simeq246$ GeV. The color convention follows that of figure~\ref{fig:1}.}
    \label{fig:2}
\end{figure}

Then, we perform an initial survey of the possible intersection points among $\lambda_e$, $\lambda_\mu$, $\lambda_\tau$, $\lambda_c$, $\lambda_b$, and $\lambda_t$ by combining the Yukawa coupling curves displayed in figure~\ref{fig:1}. The combined result is presented in figure~\ref{fig:2}.
From figure~\ref{fig:2}, one observes that the interval $800~\text{GeV}\lesssim \Lambda \lesssim 1000~\text{GeV}$ does not accommodate $\lambda_e$ and $\lambda_\mu$, while the region $1000~\text{GeV}\lesssim \Lambda \lesssim 2000~\text{GeV}$ excludes $\lambda_\tau$ and $\lambda_c$. Consequently, the viable range of $\Lambda$ in which all six Yukawa couplings can be simultaneously realized is
\begin{align}
    300~\text{GeV}\lesssim \Lambda \lesssim 700~\text{GeV}.
\end{align}
Within this interval, $\lambda_e$, $\lambda_\mu$, $\lambda_\tau$, $\lambda_c$, $\lambda_b$, and $\lambda_t$ can all be aligned within the same range. It is worth noting that, in this region, the scale-splitting corrections from Eqs.~\eqref{fm8}-\eqref{f06} can induce deviations of at most ${M_h^2 v^2}/{(4\Lambda^4)}\times 100 \simeq 0.005$--$3\%$. Therefore, the analytical approximations in Eqs.~\eqref{mm8}--\eqref{mp8} and \eqref{Ye} remain valid to at least the third decimal place.

\begin{figure}[h!]
    \centering
    \includegraphics[width=0.5\linewidth]{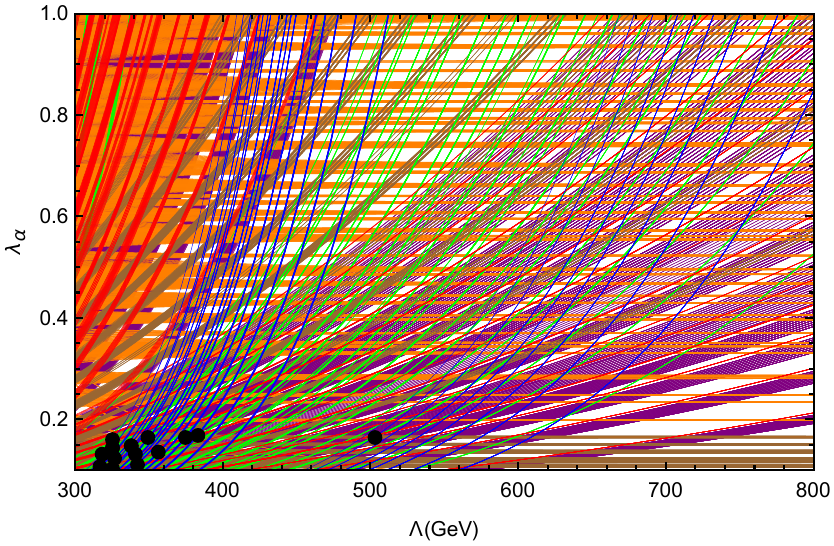}
    \caption{The dependence of Eq.~\eqref{Ye} on $\Lambda$, with $\alpha=e,\mu,\tau,b,c,t$. The scaling factors from Eqs.~\eqref{fm8}--\eqref{f06} are included. Input values are taken as $m_e\simeq0.511$ MeV, $m_\mu\simeq106$ MeV, $m_\tau\simeq1777$ MeV, $m_c=\{1.60,\dots,1.74\}$ GeV, $m_b=\{4.72,4.78,4.84\}$ GeV, $m_t=\{172.25,172.56,172.87\}$ GeV, with $M_h\simeq125$ GeV and $v\simeq246$ GeV. The color convention follows figure~\ref{fig:1}.}
    \label{fig:fig5}
\end{figure}

\begin{figure}[h!]
    \centering
    \includegraphics[width=1\linewidth]{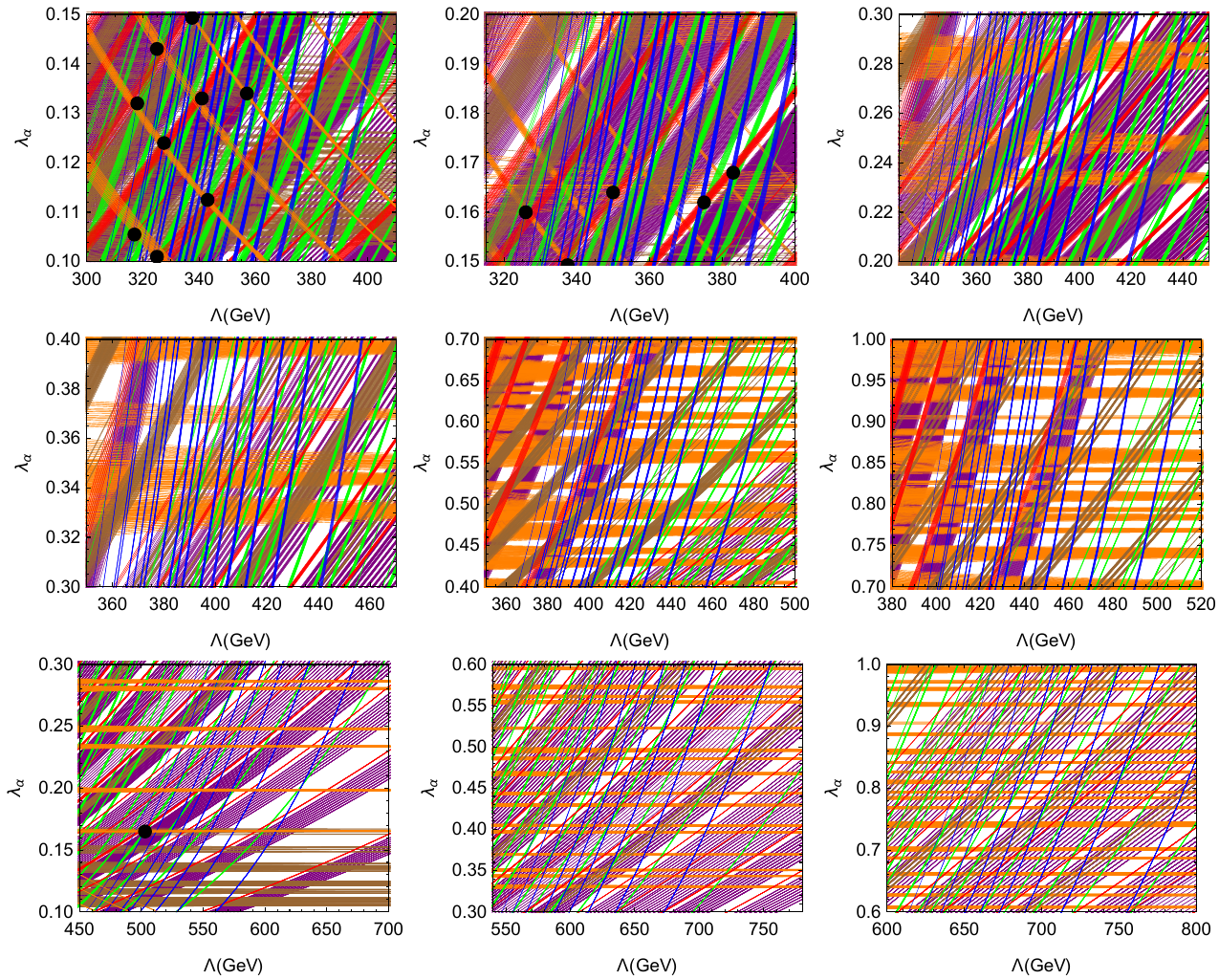}
    \caption{Magnified views of the regimes displayed in figure~\ref{fig:fig5}, showing potential intersections among different $\lambda_\alpha$. The color convention follows figure~\ref{fig:1}.}
    \label{fig:sub}
\end{figure}

Then, to explore possible intersections among the Yukawa couplings $\lambda_\alpha$, we incorporate the scale-splitting factors from Eqs.~\eqref{fm8}--\eqref{f06} together with the quoted uncertainties in the fermion masses from Eq.~\eqref{leptonmass}. The variation of $\lambda_\alpha$ as a function of $\Lambda$ in the range $300~\text{GeV}\leq \Lambda \leq 800~\text{GeV}$ is shown in figures.~\ref{fig:fig5}--\ref{fig:sub}.
Within this range, we identify 14 approximate intersection points at which subsets of Yukawa couplings approach one another. These points are denoted as $p_i=(\Lambda,\lambda_\alpha)$ and are given by
\begin{align}\label{unifiedpoints}
  &p_1\simeq (317~\text{GeV},0.106),~p_2\simeq(318~\text{GeV},0.132),\nonumber\\
  &p_3\simeq(325~\text{GeV},0.101),~p_4\simeq(325~\text{GeV},0.143),\nonumber\\
  &p_5\simeq(326~\text{GeV},0.160),~p_6\simeq(328~\text{GeV},0.124),\nonumber\\
  &p_7\simeq(338~\text{GeV},0.149),~p_8\simeq(341~\text{GeV},0.133),\nonumber\\
  &p_9\simeq(343~\text{GeV},0.113),~p_{10}\simeq(350~\text{GeV},0.164),\nonumber\\
  &p_{11}\simeq(357~\text{GeV},0.134),~p_{12}\simeq(357~\text{GeV},0.162),\nonumber\\
  &p_{13}\simeq(383~\text{GeV},0.168),~p_{14}\simeq(503~\text{GeV},0.165).
\end{align}
However, not all of these intersection points are consistent with the EW symmetry structure. By numerically evaluating Eq.~\eqref{massspectrum1}, we find that the points $(p_1,p_3,p_4,p_5,p_6,p_{10},p_{12},p_{13},p_{14})$ cannot accommodate the $SU(2)_L$ doublet structure of the top and bottom quarks. In the case $\lambda\simeq 0.16$, which corresponds to the points $\{p_{5},p_{10},p_{12},p_{13},p_{14}\}$, the resulting values of $m_b$ and $m_t$ that reproduce the observed spectrum of Eq.~\eqref{leptonmass} are
\begin{align}
    m_b&=\frac{f_{(0)}^{(1,i)}}{\sqrt{f_{L,(0)}^{(6)}f_{R,(0)}^{(6)}}}\frac{\lambda_b v}{\sqrt{2}}
    =\frac{\lambda_b v}{6\sqrt{2}}\left(1+\mathcal{O}_i(\Lambda^{-4})\right)\simeq 4.6~\text{GeV},\label{b160}
    \\
    m_t&=\frac{f_{(0)}^{(6)}}{\sqrt{f_{L,(0)}^{(1,i)}f_{R,(0)}^{(1,j)}}}\frac{\lambda_t v}{\sqrt{2}}
    =\frac{6\lambda_t v}{\sqrt{2}}\left(1+\mathcal{O}_{ij}(\Lambda^{-4})\right)\simeq 1.7\times 10^2~\text{GeV},\label{t160}
\end{align}
while for $(p_1,p_3,p_4,p_6)$, the corresponding values of $m_b$ and $m_t$ are shown below.

At $p_1\simeq(317~\text{GeV},~0.106)$:
\begin{align}
    m_b=&\{\frac{f_{Y,(0)}^{(1,i)}}{\sqrt{f_{L,(0)}^{(3,j)} f_{R,(0)}^{(5,k)}}}\frac{\lambda_b v}{\sqrt{2}},\frac{f_{Y,(0)}^{(1,i)}}{\sqrt{f_{L,(0)}^{(4,j)} f_{R,(0)}^{(4,k)}}}\frac{\lambda_b v}{\sqrt{2}}, ~(L\leftrightarrow R)\}\nonumber
     \\
     \simeq&\{\{4.52,...,4.92\},~\{4.38,...,4.76\},~(L\leftrightarrow R)\}~\text{GeV}
     \\
     m_t=&\{\frac{f_{Y,(0)}^{(2,i)}}{\sqrt{f_{L,(-4)}^{(2)}f_{R,(0)}^{(1,j)}}}\frac{\lambda_t v}{\sqrt{2}},\frac{f_{Y,(0)}^{(2,i)}}{\sqrt{f_{L,(-4)}^{(1,j)}f_{R,(0)}^{(2,k)}}}\frac{\lambda_t v}{\sqrt{2}}, \frac{f_{Y,(0)}^{(5,i)}}{\sqrt{f_{L,(-4)}^{(2)}f_{R,(0)}^{(6,j)}}}\frac{\lambda_t v}{\sqrt{2}},~(L\leftrightarrow R)\}\nonumber
     \\
     \simeq&\{\{162,...,176\},~\{163,...,175\},~\{172,...,173\},~(L\leftrightarrow R)\}~\text{GeV}
\end{align}
At $p_3\simeq(325~\text{GeV},~0.101)$:
\begin{align}
    m_b=&\{\frac{f_{Y,(0)}^{(1,i)}}{\sqrt{f_{L,(0)}^{(5,j)} f_{R,(0)}^{(3,k)}}}\frac{\lambda_b v}{\sqrt{2}}, ~(L\leftrightarrow R)\}\nonumber
     \\
     \simeq&\{\{4.36,...,4.71\},~(L\leftrightarrow R)\}~\text{GeV}
     \\
     m_t=&\{\frac{f_{Y,(-4)}^{(1,i)}}{\sqrt{f_{L,(-8)}f_{R,(-4)}^{(2)}}}\frac{\lambda_t v}{\sqrt{2}},\frac{f_{Y,(0)}^{(2,i)}}{\sqrt{f_{L,(-4)}^{(2)}f_{R,(0)}^{(1,j)}}}\frac{\lambda_t v}{\sqrt{2}},\frac{f_{Y,(0)}^{(2,i)}}{\sqrt{f_{L,(-4)}^{(1,j)}f_{R,(0)}^{(2,k)}}}\frac{\lambda_t v}{\sqrt{2}}, \frac{f_{Y,(0)}^{(4,i)}}{\sqrt{f_{L,(-4)}^{(2)}f_{R,(0)}^{(4,j)}}}\frac{\lambda_t v}{\sqrt{2}},~(L\leftrightarrow R)\}\nonumber
     \\
     \simeq&\{\{170,...,172\},~\{165,...,177\},~\{165,...,177\},~\{169,...,172\},~(L\leftrightarrow R)\}~\text{GeV}
\end{align}
At $p_4\simeq(325~\text{GeV},~0.143)$:
\begin{align}
    m_b=&\{\frac{f_{Y,(0)}^{(1,i)}}{\sqrt{f_{L,(0)}^{(5,j)} f_{R,(0)}^{(5,k)}}}\frac{\lambda_b v}{\sqrt{2}}, ~(L\leftrightarrow R)\}\nonumber
     \\
     \simeq&\{\{4.81,...,5.14\},~(L\leftrightarrow R)\}~\text{GeV}
     \\
     m_t=&\{\frac{f_{Y,(0)}^{(1,i)}}{\sqrt{f_{L,(-4)}^{(1,j)}f_{R,(0)}^{(1,k)}}}\frac{\lambda_t v}{\sqrt{2}},\frac{f_{Y,(0)}^{(2,i)}}{\sqrt{f_{L,(-4)}^{(2)}f_{R,(0)}^{(2,j)}}}\frac{\lambda_t v}{\sqrt{2}}, \frac{f_{Y,(0)}^{(2,i)}}{\sqrt{f_{L,(-4)}^{(1,j)}f_{R,(0)}^{(4,k)}}}\frac{\lambda_t v}{\sqrt{2}},~(L\leftrightarrow R)\}\nonumber
     \\
     \simeq&\{\{163,...,180\},~\{166,...,176\},~\{166,...,175\},~(L\leftrightarrow R)\}~\text{GeV}
\end{align}
At $p_6\simeq(327~\text{GeV},~0.123)$:
\begin{align}
    m_b=&\{\frac{f_{Y,(0)}^{(1,i)}}{\sqrt{f_{L,(0)}^{(5,j)} f_{R,(0)}^{(4,k)}}}\frac{\lambda_b v}{\sqrt{2}}, ~(L\leftrightarrow R)\}\nonumber
     \\
     \simeq&\{\{4.62,...,4.95\},~(L\leftrightarrow R)\}~\text{GeV}
     \\
     m_t=&\{\frac{f_{Y,(0)}^{(2,i)}}{\sqrt{f_{L,(-4)}^{(1,j)}f_{R,(0)}^{(3,k)}}}\frac{\lambda_t v}{\sqrt{2}}, \frac{f_{Y,(0)}^{(4,i)}}{\sqrt{f_{L,(-4)}^{(2)}f_{R,(0)}^{(6)}}}\frac{\lambda_t v}{\sqrt{2}},~(L\leftrightarrow R)\}\nonumber
     \\
     \simeq&\{\{167,...,177\},~\{171,...,173\},~(L\leftrightarrow R)\}~\text{GeV}\label{t123}
\end{align}
From Eqs.~\eqref{b160}--\eqref{t123}, the numerical values of $m_t$ and $m_b$ agree with the experimental masses in Eq.~\eqref{leptonmass}, but they arise from different $f_L$ factors. This implies that the $t$ and $b$ fields cannot be embedded into a single $SU(2)_L$ doublet. Consequently, the points $(p_1,p_3,p_4,p_5,p_6,p_{10},p_{12},p_{13},p_{14})$ must be excluded, since the corresponding Lagrangian cannot be consistently realized.

In addition, we find that the point $p_8\simeq(314~\text{GeV},~0.133)$ is excluded because it does not reproduce the charm quark mass within the allowed range $1.60$–$1.74$ GeV. At this point, the closest values are
\begin{align}
    \frac{f_{Y,(-4)}^{(1,i)}}{\sqrt{f_{L,(-4)}^{(2)}f_{R,(0)}^{(6)}}}\frac{\lambda_\alpha v}{\sqrt{2}}\simeq 1.77~\text{GeV},\quad
    \frac{f_{Y,(-4)}^{(1,i)}}{\sqrt{f_{L,(-4)}^{(2)}f_{R,(0)}^{(2,j)}}}\frac{\lambda_\alpha v}{\sqrt{2}}\simeq 1.51-1.56~\text{GeV}.
\end{align}
After excluding these inconsistent solutions, the viable set of intersection points reduces to only four candidates: 
\begin{align}
    \{p_2,p_7,p_9,p_{11}\}.
\end{align}
From the remaining feasible values, we obtain a mass relation that can reproduce the observed pole masses of charged leptons and heavy quarks, as given in Eq.~\eqref{leptonmass}, of the form as follows.

\noindent $\bullet$ Case $p_2$ $(\Lambda\simeq 318~\text{GeV},\lambda_\alpha\simeq 0.132)$:
\begin{align}
    m_e=&\frac{f_{Y,(-8)}}{\sqrt{f_{L,(0)}^{(6)}f_{R,(0)}^{(6)}}}\frac{\lambda_e v}{\sqrt{2}}\nonumber
    \\
    \simeq&0.511~\text{MeV}
    \\
    m_\mu=&\frac{f_{Y,(-4)}^{(1,i)}}{\sqrt{f_{L,(0)}^{(5,j)}f_{R,(0)}^{(5,k)}}}\frac{\lambda_\mu v}{\sqrt{2}}\nonumber
    \\
    \simeq&\{105,...,107\}~\text{MeV}
    \\
    m_\tau=&\{\frac{f_{Y,(-4)}^{(1,i)}}{\sqrt{f_{L,(-4)}^{(2)}f_{R,(0)}^{(2,j)}}}\frac{\lambda_\tau v}{\sqrt{2}},~(L\leftrightarrow R)\}\nonumber
    \\
    \simeq&\{\{1.72,...,1.78\},~(L\leftrightarrow R)\}~\text{GeV}
     \\
    m_c=&\{\frac{f_{Y,(-4)}^{(1,i)}}{\sqrt{f_{L,(-4)}^{(2)}f_{R,(0)}^{(2,j)}}}\frac{\lambda_c v}{\sqrt{2}},~\frac{f_{Y,(-4)}^{(1,i)}}{\sqrt{f_{L,(-4)}^{(1,j)}f_{R,(0)}^{(4,k)}}}\frac{\lambda_c v}{\sqrt{2}},~\frac{f_{Y,(-8)}}{\sqrt{f_{L,(-4)}^{(1,i)}f_{R,(-8)}}}\frac{\lambda_c v}{\sqrt{2}},~(L\leftrightarrow R)\}\nonumber
    \\
    \simeq&\{\{1.72,...,1.78\},~\{1.73,...,1.76\},~\{1.74,...,1.75\},~(L\leftrightarrow R)\}~\text{GeV}
    \\
    m_b=&\{\frac{f_{Y,(0)}^{(1,i)}}{\sqrt{f_{L,(0)}^{(6)}f_{R,(0)}^{(4,j)}}}\frac{\lambda_b v}{\sqrt{2}},~\frac{f_{Y,(0)}^{(1,i)}}{\sqrt{f_{L,(0)}^{(5,j)}f_{R,(0)}^{(5,k)}}}\frac{\lambda_b v}{\sqrt{2}},~(L\leftrightarrow R)\}\nonumber
    \\
    \simeq&\{\{4.51,...,4.86\},~\{4.51,...,4.86\},~(L\leftrightarrow R)\}~\text{GeV}
    \\
    m_t=&\{\frac{f_{Y,(0)}^{(2,i)}}{\sqrt{f_{L,(-4)}^{(1,j)}f_{R,(0)}^{(3,k)}}}\frac{\lambda_t v}{\sqrt{2}},~\frac{f_{Y,(0)}^{(4,i)}}{\sqrt{f_{L,(-4)}^{(2)}f_{R,(0)}^{(6)}}}\frac{\lambda_t v}{\sqrt{2}},~(L\leftrightarrow R)\}\nonumber
    \\
    \simeq&\{\{169,...,180\},~\{173,...,175\},~(L\leftrightarrow R)\}~\text{GeV}
\end{align}
According to the SU(2) constraint, the configuration of $t$ and $b$ mass is
\begin{align}
    m_t=\frac{f_{Y,(0)}^{(4,1)}}{\sqrt{f_{R,(-4)}^{(2)}f_{L,(0)}^{(6)}}}\frac{\lambda_t v}{\sqrt{2}}, ~m_b=\frac{f_{Y,(0)}^{(1,i)}}{\sqrt{f_{L,(0)}^{(6)}f_{R,(0)}^{(4,j)}}}\frac{\lambda_b v}{\sqrt{2}}.
\end{align}
$\bullet$ Case $p_7\simeq(337~\text{GeV},~0.149)$:
\begin{align}
    m_e=&\{\frac{f_{Y,(-8)}}{\sqrt{f_{L,(0)}^{(3,i)}f_{R,(0)}^{(6)}}}\frac{\lambda_e v}{\sqrt{2}},~(L\leftrightarrow R)\}\nonumber
    \\
    \simeq&\{\{0.511,...,0.515\},~(L\leftrightarrow R)\}~\text{MeV}
    \\
    m_\mu=&\{\frac{f_{Y,(-8)}}{\sqrt{f_{L,(-8)}f_{R,(0)}^{(4,i)}}}\frac{\lambda_\mu v}{\sqrt{2}},\frac{f_{Y,(-4)}^{(1,i)}}{\sqrt{f_{L,(0)}^{(5.j)}f_{R,(0)}^{(5,k)}}}\frac{\lambda_\mu v}{\sqrt{2}},~(L\leftrightarrow R)\}\nonumber
    \\
    \simeq&\{106,~\{105,...,107\},~(L\leftrightarrow R)\}~\text{MeV}
    \\
    m_\tau=&\{\frac{f_{Y,(-4)}^{(1,i)}}{\sqrt{f_{L,(-4)}^{(2)}f_{R,(0)}^{(2,j)}}}\frac{\lambda_\tau v}{\sqrt{2}},~\frac{f_{Y,(-4)}^{(1,i)}}{\sqrt{f_{L,(-4)}^{(1,j)}f_{R,(0)}^{(4,k)}}}\frac{\lambda_\tau v}{\sqrt{2}},~(L\leftrightarrow R)\}\nonumber
    \\
    \simeq&\{\{1.72,...,1.78\},~\{1.74,...,1.77\},~(L\leftrightarrow R)\}~\text{GeV}
      \end{align}
    \begin{align}
    m_c=&\{\frac{f_{Y,(-4)}^{(1,i)}}{\sqrt{f_{L,(-4)}^{(2)}f_{R,(0)}^{(2,j)}}}\frac{\lambda_\tau v}{\sqrt{2}},~\frac{f_{Y,(-4)}^{(1,i)}}{\sqrt{f_{L,(-4)}^{(1,j)}f_{R,(0)}^{(4,k)}}}\frac{\lambda_\tau v}{\sqrt{2}},~(L\leftrightarrow R)\}\nonumber
    \\
    \simeq&\{\{1.72,...,1.78\},~\{1.74,...,1.77\},~(L\leftrightarrow R)\}~\text{GeV}
   \\
    m_b=&\{\frac{f_{Y,(0)}^{(1,i)}}{\sqrt{f_{L,(0)}^{(6)}f_{R,(0)}^{(5,j)}}}\frac{\lambda_b v}{\sqrt{2}},~(L\leftrightarrow R)\}\nonumber
    \\
    \simeq&\{\{4.60,...,4.87\},~(L\leftrightarrow R)\}~\text{GeV}
    \\
    m_t=&\{\frac{f_{Y,(0)}^{(2,i)}}{\sqrt{f_{L,(-4)}^{(1,j)}f_{R,(0)}^{(5,k)}}}\frac{\lambda_t v}{\sqrt{2}},~(L\leftrightarrow R)\}\nonumber
    \\
    \simeq&\{\{169,...,177\},~(L\leftrightarrow R)\}~\text{GeV}
\end{align}
According to the SU(2) constraint, the configuration of $t$ and $b$ mass is
\begin{align}
    m_t=\frac{f_{Y,(0)}^{(2,i)}}{\sqrt{f_{R,(-4)}^{(1,j)}f_{L,(0)}^{(5,k)}}}\frac{\lambda_t v}{\sqrt{2}}, ~m_b=\frac{f_{Y,(0)}^{(1,l)}}{\sqrt{f_{R,(0)}^{(6)}f_{L,(0)}^{(5,k)}}}\frac{\lambda_b v}{\sqrt{2}}.
\end{align}
$\bullet$ Case $p_9\simeq(343~\text{GeV},~0.113)$:
\begin{align}
    m_e=&\{\frac{f_{Y,(-8)}}{\sqrt{f_{L,(0)}^{(2,i)}f_{R,(0)}^{(4,j)}}}\frac{\lambda_e v}{\sqrt{2}},~(L\leftrightarrow R)\}\nonumber
    \\
    \simeq&\{\{0.501,...,0.512\},~(L\leftrightarrow R)\}~\text{MeV}
    \\
    m_\mu=&\{\frac{f_{Y,(-4)}^{(1,i)}}{\sqrt{f_{L,(0)}^{(5,j)}f_{R,(0)}^{(2,k)}}}\frac{\lambda_\mu v}{\sqrt{2}},~(L\leftrightarrow R)\}\nonumber
    \\
    \simeq&\{\{105,...,108\},~(L\leftrightarrow R)\}~\text{MeV}
    \\
    m_\tau=&\{\frac{f_{Y,(-4)}^{(1,i)}}{\sqrt{f_{L,(-4)}^{(1,j)}f_{R,(0)}^{(2,k)}}}\frac{\lambda_\tau v}{\sqrt{2}},~(L\leftrightarrow R)\}\nonumber
    \\
    \simeq&\{\{1.79,...,1.84\},~(L\leftrightarrow R)\}~\text{GeV}
     \\
    m_c=&\{\frac{f_{Y,(-4)}^{(2)}}{\sqrt{f_{L,(-4)}^{(2)}f_{R,(0)}^{(5,j)}}}\frac{\lambda_\tau v}{\sqrt{2}},~(L\leftrightarrow R)\}\nonumber
    \\
    \simeq&\{\{1.62,...,1.63\},~(L\leftrightarrow R)\}~\text{GeV}
    \\
    m_b=&\{\frac{f_{Y,(0)}^{(1,i)}}{\sqrt{f_{L,(0)}^{(6)}f_{R,(0)}^{(3,j)}}}\frac{\lambda_b v}{\sqrt{2}},\frac{f_{Y,(0)}^{(1,i)}}{\sqrt{f_{L,(0)}^{(4,j)}f_{R,(0)}^{(4,k)}}}\frac{\lambda_b v}{\sqrt{2}},~(L\leftrightarrow R)\}\nonumber
    \\
    \simeq&\{\{4.50,...,4.77\},~\{4.77,...,5.06\},~(L\leftrightarrow R)\}~\text{GeV}
    \\
    m_t=&\{\frac{f_{Y,(0)}^{(2,i)}}{\sqrt{f_{L,(-4)}^{(1,j)}f_{R,(0)}^{(3,k)}}}\frac{\lambda_t v}{\sqrt{2}},\frac{f_{Y,(0)}^{(4,i)}}{\sqrt{f_{L,(-4)}^{(2)}f_{R,(0)}^{(6)}}}\frac{\lambda_t v}{\sqrt{2}},~(L\leftrightarrow R)\}\nonumber
    \\
    \simeq&\{\{170,...,178\},~\{173,...,174\},~(L\leftrightarrow R)\}~\text{GeV}
\end{align}
According to the SU(2) constraint, the configurations of $t$ and $b$ mass are
\begin{align}
    m_t=\frac{f_{Y,(0)}^{(2,i)}}{\sqrt{f_{R,(-4)}^{(1,j)}f_{L,(0)}^{(3,k)}}}\frac{\lambda_t v}{\sqrt{2}}, ~m_b=\frac{f_{Y,(0)}^{(1,l)}}{\sqrt{f_{R,(0)}^{(6)}f_{L,(0)}^{(3,k)}}}\frac{\lambda_b v}{\sqrt{2}},
    \\
    m_t=\frac{f_{Y,(0)}^{(4,i)}}{\sqrt{f_{R,(-4)}^{(2)}f_{L,(0)}^{(6)}}}\frac{\lambda_t v}{\sqrt{2}},~m_b=\frac{f_{Y,(0)}^{(1,j)}}{\sqrt{f_{R,(0)}^{(3,k)}f_{L,(0)}^{(6)}}}\frac{\lambda_b v}{\sqrt{2}}.
\end{align}
$\bullet$ Case $p_{11}\simeq(357~\text{GeV},~0.134)$:
\begin{align}
    m_e=&\{\frac{f_{Y,(-8)}}{\sqrt{f_{L,(0)}^{(2,i)}f_{R,(0)}^{(3,j)}}}\frac{\lambda_e v}{\sqrt{2}},~(L\leftrightarrow R)\}\nonumber
    \\
    \simeq&\{\{0.510,...,0.521\},~(L\leftrightarrow R)\}~\text{MeV}
    \\
    m_\mu=&\{\frac{f_{Y,(-4)}^{(1,i)}}{\sqrt{f_{L,(0)}^{(5,j)}f_{R,(0)}^{(2,k)}}}\frac{\lambda_\mu v}{\sqrt{2}},~(L\leftrightarrow R)\}\nonumber
    \\
    \simeq&\{\{106,...,109\},~(L\leftrightarrow R)\}~\text{MeV}
    \\
    m_\tau=&\{\frac{f_{Y,(-4)}^{(2)}}{\sqrt{f_{L,(-4)}^{(2)}f_{R,(0)}^{(5,i)}}}\frac{\lambda_\tau v}{\sqrt{2}},~(L\leftrightarrow R)\}\nonumber
    \\
    \simeq&\{1.78,~(L\leftrightarrow R)\}~\text{GeV}
     \\
    m_c=&\{\frac{f_{Y,(-4)}^{(2)}}{\sqrt{f_{L,(-4)}^{(2)}f_{R,(0)}^{(6)}}}\frac{\lambda_\tau v}{\sqrt{2}},~(L\leftrightarrow R)\}\nonumber
    \\
    \simeq&\{1.62,~(L\leftrightarrow R)\}~\text{GeV}
    \\
    m_b=&\{\frac{f_{Y,(0)}^{(1,i)}}{\sqrt{f_{L,(0)}^{(6)}f_{R,(0)}^{(4,j)}}}\frac{\lambda_b v}{\sqrt{2}},\frac{f_{Y,(0)}^{(1,i)}}{\sqrt{f_{L,(0)}^{(5,j)}f_{R,(0)}^{(5,k)}}}\frac{\lambda_b v}{\sqrt{2}},~(L\leftrightarrow R)\}\nonumber
    \\
    \simeq&\{\{4.65,...,4.87\},~\{4.55,...,4.77\},~(L\leftrightarrow R)\}~\text{GeV}
    \\
    m_t=&\{\frac{f_{Y,(0)}^{(2,i)}}{\sqrt{f_{L,(-4)}^{(1,j)}f_{R,(0)}^{(5,k)}}}\frac{\lambda_t v}{\sqrt{2}},~(L\leftrightarrow R)\}\nonumber
    \\
    \simeq&\{\{170,...,176\},~(L\leftrightarrow R)\}~\text{GeV}
\end{align}
The configuration of $t$ and $b$ mass that satisfy  the SU(2) constraint is
\begin{align}
    m_t=\frac{f_{Y,(0)}^{(2,i)}}{\sqrt{f_{R,(-4)}^{(1,j)}f_{L,(0)}^{(5,k)}}}\frac{\lambda_t v}{\sqrt{2}},~m_b=\frac{f_{Y,(0)}^{(1,l)}}{\sqrt{f_{L,(0)}^{(5,k)}f_{R,(0)}^{(5,m)}}}\frac{\lambda_b v}{\sqrt{2}}.
\end{align}
Then, we turn to the discussion of the results.  

Among the four feasible points, there are two particularly noteworthy points, which are
\begin{align}
p_2\simeq(318~\text{GeV},~0.132), \qquad p_{11}\simeq(357~\text{GeV},~0.134).
\end{align}
Within these points, Yukawa couplings of the charged leptons and heavy quarks approximately converge to an universal value that is numerically close to the Higgs self-coupling,
\begin{align}   \lambda_e\simeq\lambda_\mu\simeq\lambda_\tau\simeq\lambda_c\simeq\lambda_b\simeq\lambda_t\simeq\lambda
    \simeq \frac{M_h^2}{2v^2}\simeq 0.13.
\end{align}
At these points, the number of independent dimensionless parameters in the Yukawa sector and the Higgs sector is effectively minimized. Although the framework introduces one additional dimensionful scale $\Lambda$, up to seven Yukawa couplings are possibly reduced to a single effective parameter.

In addition, the point $p_2$ exhibits a further notable feature: the electron mass emerges as the lightest fermion mass,
\begin{align}
    M_e=\frac{f_{Y,(-8)}}{\sqrt{f^{Largest}_{L,(0)}f^{Largest}_{R,(0)}}}\frac{\lambda_e v}{\sqrt{2}}=\frac{f_{Y,(-8)}}{\sqrt{f^{(6)}_{L,(0)}f^{(6)}_{R,(0)}}}\frac{\lambda_e v}{\sqrt{2}}\simeq\frac{\lambda_eM_h^4 v^5}{384\sqrt{2}\Lambda^8}\simeq0.511~\text{MeV}
\end{align}
generated through the Higgs mechanism, consistent with the observed ordering of the charged-lepton spectrum. This coincidence may indicate that $p_2$ provides a more natural realization of the fermion mass hierarchy compared to other solutions.

Interestingly, the point $p_{11}$ is no less significant than $p_2$. 
At this point, the scaling factor in our framework can naturally approach to the Cabibbo angle, $(\theta_c)$ \cite{ParticleDataGroup:2024cfk},
\begin{align}
    \frac{M_h v}{\Lambda}\sim \theta_c\sim 0.2.
\end{align}
This behavior is analogous to the Froggatt--Nielsen mechanism, where a comparable suppression 
originates from the ratio between the flavon vacuum expectation value and the effective field theory scale,
$\langle \phi \rangle / \Lambda_{\mathrm{EFT}}\sim 0.2$, which is also expected to lie within this range.
Furthermore, the numerical coincidence between the scaling factor and $\theta_c$ suggests that the present framework 
may provide a new perspective on flavor structures, potentially linking the Higgs sector with the observed pattern of quark mixing. In addition, at the point $p_{11}$, an intriguing feature arises in the context of symmetry extensions. We observe that the top quark, bottom quark, and $\tau$ lepton share an identical left-handed profile factor $f_L$. This observation suggests the possibility of embedding these three fermion fields within a common multiplet, 
\begin{align}
    \Psi_L=\begin{pmatrix}
        b\\ t\\\tau
    \end{pmatrix}_L
\end{align}
as naturally realized in extended gauge structures such as the chiral $SU(4)$ model \cite{Long:2016lmj, PhysRevD.99.015029}, known as 3-4-1 model.

In summary, within the multiplicative Higgs Lagrangian framework, the observed masses of charged leptons and heavy quarks can be consistently accommodated without invoking widely separated scale of Yukawa couplings, as expressed in Eqs.~\eqref{cone}–\eqref{cont}. Furthermore, this formulation naturally admits the possibility of equal Yukawa couplings between the charged leptons and heavy quarks under specific parameter configurations.

\section{Discussions}\label{sec:discussion}
We now turn to further implications of the multiplicative Higgs Lagrangian and outline possible directions for future work.  

\subsection{Predictive Signatures}
In this section, we investigate the potential phenomenological signatures of the multiplicative Higgs framework. If the observed charged-lepton and heavy-quark masses are reproduced at one of the representative parameter points 
$\{p_2, p_7, p_9, p_{11}\}$, the model predicts a characteristic pattern of the Wilson coefficients in 
Eqs.~\eqref{L5}--\eqref{L6} at the matching scale $\Lambda_\text{EFT} = 246~\text{GeV}$ \cite{Ellis:2018gqa}, as summarized below.

\noindent For $p_2=(\Lambda\simeq318~\text{GeV},\lambda_\alpha=0.132)$: 
\begin{align}
   &c_5=6.90\times10^{-5},~c_{3hWW}=4.01\times 10^{-4} g,~c_{3hZZ}=2.00\times10^{-4}(g^2+g'^2),\nonumber
   \\
   &c_{HD}=2.67\times 10^{-4},~c_6=5.75\times 10^{-6},~c_{4hWW}=1.67\times 10^{-4}g^2,~c_{4hZZ}= 8.35\times 10^{-5}(g^2+g'^2).
\end{align}
For $p_7=(\Lambda\simeq337~\text{GeV},\lambda_\alpha=0.149)$
\begin{align}
   &c_5=4.34\times10^{-5},~c_{3hWW}=2.52\times 10^{-4} g,~c_{3hZZ}=1.26\times10^{-4}(g^2+g'^2),\nonumber
   \\
   &c_{HD}=1.68\times 10^{-4},~c_6=3.61\times 10^{-6},~c_{4hWW}=1.05\times 10^{-4}g^2,~c_{4hZZ}= 5.25\times 10^{-5}(g^2+g'^2).
\end{align}
For $p_9=(\Lambda\simeq343~\text{GeV},\lambda_\alpha=0.113)$
\begin{align}
   &c_5=3.77\times10^{-5},~c_{3hWW}=2.19\times 10^{-4} g,~c_{3hZZ}=1.09\times10^{-4}(g^2+g'^2),\nonumber
   \\
   &c_{HD}=1.46\times 10^{-4},~c_6=3.14\times 10^{-6},~c_{4hWW}=9.11\times 10^{-5}g^2,~c_{4hZZ}= 4.56\times 10^{-5}(g^2+g'^2).
\end{align}
For $p_{11}=(\Lambda\simeq357~\text{GeV},\lambda_\alpha=0.134)$
\begin{align}
   &c_5=2.73\times10^{-5},~c_{3hWW}=1.59\times 10^{-4} g,~c_{3hZZ}=7.94\times10^{-5}(g^2+g'^2),\nonumber
   \\
   &c_{HD}=1.06\times 10^{-4},~c_6=2.28\times 10^{-6},~c_{4hWW}=6.62\times 10^{-5}g^2,~c_{4hZZ}= 3.31\times 10^{-5}(g^2+g'^2).
\end{align}
The corresponding Wilson coefficients evaluated at $\Lambda_{\text{EFT}} = 1~\text{TeV}$ are provided in appendix~\ref{secWilson}. To achieve $c_i \simeq 1$, the corresponding tree-level matching requires an energy scale of $\Lambda_{\text{EFT}} \simeq 30~\text{TeV}$. These coefficients may exhibit distinctive patterns that could be probed in future Higgs precision measurements~\cite{Observation10, Observation11, Mastandrea:2024irf, Benedikt:2022kan}, thereby providing a potential avenue to experimentally falsify the framework.

\subsection{UV behavior in the large background field limit}
It is instructive to compare the UV properties of our Higgs framework with those of the Standard Model (SM) in the regime of a large classical background field $\phi_b$. In the SM, the background-dependent Higgs self-couplings take the form
\begin{align}
    \lambda_{3,\text{SM}}(\phi_b)=\frac{3M_h^2 \phi_b}{v^2}, 
    \qquad 
    \lambda_{4,\text{SM}}(\phi_b)=\frac{3M_h^2}{v^2}.
\end{align}
The trilinear self-coupling grows linearly with $\phi_b$, implying that perturbative unitarity in scattering amplitudes eventually breaks down at large field values.  

\begin{figure}[h]
    \centering
    \includegraphics[width=0.5\linewidth]{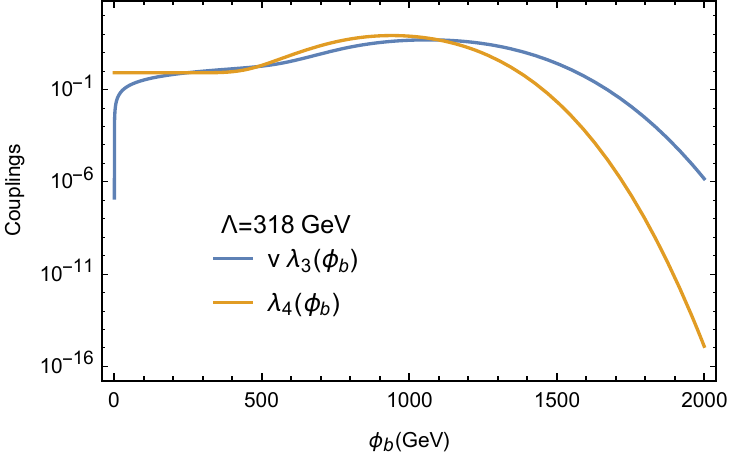}
    \caption{Background-field dependence of the trilinear and quartic Higgs couplings in the multiplicative Lagrangian, for $M_h=125~\text{GeV}$, $v=246~\text{GeV}$, and $\Lambda=318~\text{GeV}$.}
    \label{fig:bd}
\end{figure}

In contrast, in our model the effective couplings $\lambda_3(\phi_b)$ and $\lambda_4(\phi_b)$ exhibit a qualitatively different scaling, as shown in figure~\ref{fig:bd}. The expression of $\lambda_3(\phi_b)$ and $\lambda_4(\phi_b)$ predicted from this model are shown in appendix C. In the range $0<\phi_b<v$, the couplings prefers the SM prediction, $\lambda_{3}(\phi_b)\simeq \lambda_{3,\text{SM}}(\phi_b)$ and $\lambda_{4}(\phi_b)\simeq \lambda_{4,\text{SM}}(\phi_b)$. In the limit $\phi_b\gg v$,  both couplings asymptotically decrease toward very small values as $\phi_b$ increases,
\begin{align}
    \lambda_3(\phi_b)\simeq \frac{\phi _b^9 M_h^6 e^{\frac{\phi _b^4 M_h^2}{8 \Lambda ^4 v^2}}}{16 \Lambda ^8 v^6 \cosh ^{\frac{3}{2}}\left(\frac{\phi _b^4 M_h^2}{8 \Lambda ^4
   v^2}\right)},~\lambda_4(\phi_b)\simeq \frac{\phi _b^{12} M_h^8 e^{-\frac{\phi _b^4 M_h^2}{8 \Lambda ^4 v^2}}}{8 \Lambda ^{12} v^8},
\end{align}
in stark contrast to the linear growth in the SM, illustrated in figure~\ref{fig:bd}. This behavior suggests that perturbativity of the scalar sector can remain valid even in the large-field regime.  

\begin{figure}[h]
    \centering
    \includegraphics[width=0.5\linewidth]{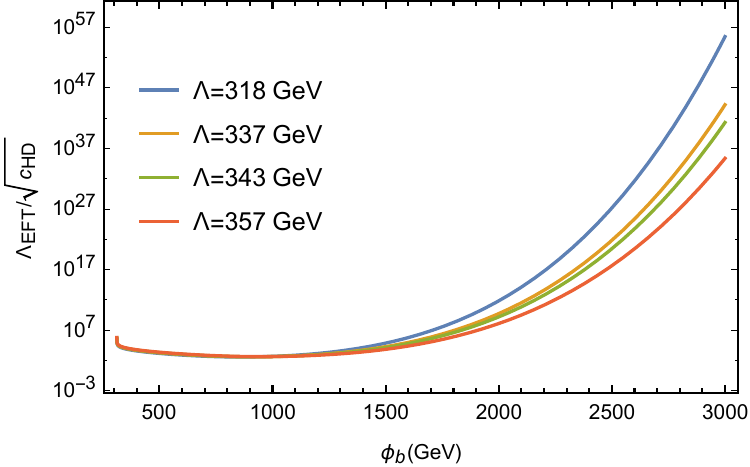}
    \caption{Background-field dependence of the effective cutoff $\Lambda_\text{EFT}/\sqrt{c_{HD}}$ for representative values of $\Lambda$. The growth beyond the Planck scale suggests strong suppression of higher-dimensional operators in the large-field regime.}
    \label{fig:cutoffb}
\end{figure}

Moreover, the UV cutoff scale of the low-energy effective theory, estimated by the most dangerous operator $\Lambda_\text{EFT}/\sqrt{c_{HD}}$, also acquires a strong background dependence. As shown in figure~\ref{fig:cutoffb}, the UV cutoff grows rapidly with $\phi_b$ and can exceed the Planck scale for $\phi_b$ in the TeV range. This indicates that higher-dimensional operators are highly suppressed below the Planck scale. Therefore, below the Planck scale, the Higgs Lagrangian in the large-field regime can be approximated by that of a free massive scalar field, 
\begin{align}
    \mathcal{L}_{\text{Higgs}}\simeq \frac{\partial_\mu h\partial^\mu h}{2}-\frac{M_h(\phi_b)^2h^2}{2}
\end{align}
with the cubic background dependent effective mass 
as 
\begin{align}
  M_h(\phi_b)^2=  -\frac{4 \Lambda ^4 \left(\mu ^2-3 \lambda  \phi _b^2\right)+\phi _b^2 \left(\mu ^2-\lambda  \phi _b^2\right){}^2 \tanh \left(\frac{2 \mu ^2 \phi
   _b^2-\lambda  \phi _b^4}{16 \Lambda ^4}\right)}{16 \Lambda ^4}\simeq \frac{\lambda ^2 \phi _b^6}{16 \Lambda ^4}.
\end{align}
Consequently, the one-loop effective potential \cite{PhysRevD.7.1888, Markkanen:2018bfx, PhysRevD.96.096005}
\begin{align}
    U_\text{1-loop}=\sum_i (-1)^{2s_i}\frac{n_d^i}{64\pi^2}m_i^4(\phi_b)\log\frac{m_i^2}{\tilde{\mu}^2}
\end{align}
can possibly be expected to be stabilized by the cubic Higgs mass term, as the background dependent masses in the SM exhibit a linear dependence on the Higgs field $m_\text{SM}(\phi_b)\propto \phi_b$. Here, $n_d$ is 12 for color Dirac particle and 1 for neutral scale, $(-1)^{2s_i}$ is -1 for fermion and 1 for boson and $\tilde{\mu}$ is renormalized scale. 

An open question is whether this asymptotic behavior can be extended to gauge and Yukawa interactions. If similar suppression can emerges across all couplings, the multiplicative Higgs framework might allow a consistent perturbative description in large-field regimes. This could have implications for scenarios where the Higgs background plays a dynamical role in the early universe, including but not limited to Higgs-inflation–type models
\cite{HiggsInflation,CiteTheActionOfHiggsInflation,CiteTheActionOfHiggsInflation2,CiteTheActionOfHiggsInflation3}. A more detailed analysis of radiative corrections and renormalization group running in the presence of the multiplicative structure would be required to clarify this possibility.  

\section{Conclusions}\label{conclusion}
We have investigated a multiplicative Higgs Lagrangian that reformulates the fermion mass hierarchy problem in terms of discrete Lagrangian configurations rather than tuning continuous value of Yukawa couplings. 
This framework admits candidate points where the charged lepton and heavy quark Yukawa couplings are approximately equal with the Higgs self-coupling, leading to a striking reduction in the number of free parameters. 
In addition, the model exhibits improved ultraviolet behavior, with background-dependent couplings that remain perturbative even in the large-field regime. 
These results suggest that the multiplicative Higgs construction may provide a useful alternative perspective on fermion mass generation and motivate further exploration of its embedding into a UV-complete theory.

\section*{Acknowledgments}
This research has received funding support from the NSFR via the Program Management Unit for Human Resource and Institutional Development Research and Innovation [grant number B13F670067]. This work was financially supported by King Mongkut's Institute of Technology Ladkrabang [KREF186713].
\appendix
\section{choices of epsilons and scaling factor}\label{AppA}
$\bullet$ case $f_{(-8)}$: (-1,0,-1,0,-1),~(-1,0,-1,1,0),~(-1,1,0,0,-1), (-1,1,0,1,0)

$\bullet$ case $f_{(-4)}^{(1,1)}$: (1,-1,1,1,1), (1,-1,1,-1,-1), (1,-1,1,0,0)

$\bullet$ case $f_{(-4)}^{(1,2)}$: (0,0,1,1,0), (0,0,1,0,-1), (0,-1,0,1,0), (0,-1,0,0,-1)

$\bullet$ case $f_{(-4)}^{(1,3)}$: (-1,0,0,1,-1), (-1,1,1,1,-1), (-1,-1,-1,1,-1)

$\bullet$ case $ f_{(-4)}^{(2)}$: (0,-1,1,1,-1)

$\bullet$ case $f_{(0)}^{(1,1)}$: (0,1,-1,0,1), (1,1,0,-1,1), (1,0,-1,-1,1), (0,1,-1,-1,0)

$\bullet$ case $f_{(0)}^{(1,2)}$: (-1,1,-1,0,-1), (-1,1,-1,1,0), (0,0,-1,0,0), (0,0,-1,-1,-1), (0,0,-1,1,1), (0,1,0,-1,-1), (0,1,0,1,1), (1,1,1,0,1), (1,1,1,-1,0), (1,0,0,0,1), (1,0,0,-1,0), (1,-1,-1,0,1), (1,-1,-1,-1,0), (0,1,0,0,0)

$\bullet$ case $f_{(0)}^{(1,3)}$: (-1,0,-1,1,-1), (-1,1,0,1,-1), (0,-1,-1,1,0), (0,-1,-1,0,-1), (0,0,0,0,-1),(0,1,1,0,-1), (1,0,1,1,1), (1,0,1,0,0), (1,0,1,-1,-1), (1,-1,0,1,1), (1,-1,0,-1,-1), (0,1,1,1,0), (1,-1,0,0,0), (0,0,0,1,0)

$\bullet$ case $f_{(0)}^{(1,4)}$: (0,-1,0,1,-1), (1,-1,1,1,0), (1,-1,1,0,-1), (0,0,1,1,0,-1)

$\bullet$ case $f_{(0)}^{(2,1)}$: (1,1,-1,-1,1)

$\bullet$ case $ f_{(0)}^{(2,2)}$: (1,0,-1,0,1), (0,1,-1,-1,-1), (0,1,-1,0,0), (0,1,-1,1,1), (1,0,-1,-1,0), (1,1,0,0,1), (1,1,0,-1,0)

$\bullet$ case $ f_{(0)}^{(2,3)}$: (1,-1,-1,0,0), (1,-1,-1,1,1), (1,-1,-1,-1,-1)$_{X}$, (1,0,0,0,0), (1,0,0,-1,-1), (1,1,1,0,0), (1,1,1,1,1), (1,1,1,-1,-1), (1,0,0,1,1)

$\bullet$ case $f_{(0)}^{(2,4)}$: (0,0,-1,0,-1), (0,0,-1,1,0), (0,1,0,0,-1), (0,1,0,1,0)

$\bullet$ case $f_{(0)}^{(2,5)}$: (-1,1,-1,1,-1)

$\bullet$ case $f_{(0)}^{(2,6)}$: (0,-1,-1,1,-1), (0,0,0,1,-1), (1,-1,0,0,-1), (1,-1,0,1,0), (0,1,1,1,-1), (1,0,1,1,0), (1,0,1,0,-1)

$\bullet$ case $f_{(0)}^{(2,7)}$: (1,-1,1,1,-1)

$\bullet$ case $f_{(0)}^{(3,1)}$: (1,1,-1,-1,0), (1,1,-1,0,1)

$\bullet$ case $f_{(0)}^{(3,2)}$: (0,1,-1,0,-1), (0,1,-1,1,0), (1,1,0,1,1), (1,1,0,0,0), (1,1,0,-1,-1), (1,0,-1,1,1), (1,0,-1,0,0), (1,0,-1,-1,-1)

$\bullet$ case $ f_{(0)}^{(3,3)}$: (0,0,-1,1,-1), (0,1,0,1,-1), (1,-1,-1,0,-1), (1,1,1,0,-1), (1,1,1,1,0), (1,0,0,1,0), (1,0,0,0,-1), (1,-1,-1,1,0)

$\bullet$ case $ f_{(0)}^{(3,4)}$: (1,0,1,1,-1), (1,-1,0,1,-1)

$\bullet$ case $  f_{(0)}^{(4,1)}$: (1,1,-1,1,1), (1,1,-1,0,0), (1,1,-1,-1,-1)

$\bullet$ case $ f_{(0)}^{(4,2)}$: (1,1,0,1,0), (1,1,0,0,-1), (1,0,-1,1,0), (1,0,-1,0,-1)

$\bullet$ case $ f_{(0)}^{(4,3)}$:(0,1,-1,1,-1)

$\bullet$ case $ f_{(0)}^{(4,4)}$: (1,1,1,1,-1), (1,0,0,1,-1), (1,-1,-1,1,-1)

$\bullet$ case $ f_{(0)}^{(5,1)}$: (1,1,-1,1,0), (1,1,-1,0,-1)

$\bullet$ case $ f_{(0)}^{(5,2)}$: (1,0,-1,1,-1), (1,1,0,1,-1)

$\bullet$ case $f_{(0)}^{(6)}$: (1,1,-1,1,-1)

\section{Wilson's coefficients}\label{secWilson}
At $\Lambda_\text{EFT}=1000~\text{GeV}$, this model predict the Wilson coefficients for each intersection point of Yukawa coupling as follows.

\noindent For $p_2=(\Lambda\simeq318~\text{GeV},\lambda_\alpha=0.132)$: 
\begin{align}
    &c_5=2.80\times 10^{-4},~c_{3hWW}=1.63\times 10^{-3}g^2,~c_{3hZZ}=8.15\times 10^{-4}(g^2+g'^2),\nonumber
    \\
    &c_{HD}=4.42\times 10^{-3},~c_{6}=9.50\times 10^{-5},~c_{4hWW}=2.76\times 10^{-3} g^2,~c_{4hZZ}=1.38\times 10^{-3} (g^2+g'^2).
\end{align}
For $p_7=(\Lambda\simeq337~\text{GeV},\lambda_\alpha=0.149)$: 
\begin{align}
    &c_5=1.76\times 10^{-4},~c_{3hWW}=1.02\times 10^{-3}g^2,~c_{3hZZ}=5.12\times 10^{-4}(g^2+g'^2),\nonumber
    \\
    &c_{HD}=2.78\times 10^{-3},~c_{6}=5.97\times 10^{-5},~c_{4hWW}=1.73\times 10^{-3} g^2,~c_{4hZZ}=8.67\times 10^{-4} (g^2+g'^2).
\end{align}
For $p_9=(\Lambda\simeq343~\text{GeV},\lambda_\alpha=0.113)$: 
\begin{align}
    &c_5=1.53\times 10^{-4},~c_{3hWW}=8.89\times 10^{-4}g^2,~c_{3hZZ}=4.45\times 10^{-4}(g^2+g'^2),\nonumber
    \\
    &c_{HD}=2.41\times 10^{-3},~c_{6}=5.18\times 10^{-5},~c_{4hWW}=1.51\times 10^{-3} g^2,~c_{4hZZ}=7.53\times 10^{-4} (g^2+g'^2).
\end{align}
For $p_{11}=(\Lambda\simeq357~\text{GeV},\lambda_\alpha=0.134)$: 
\begin{align}
    &c_5=1.11\times 10^{-4},~c_{3hWW}=6.46\times 10^{-4}g^2,~c_{3hZZ}=3.23\times 10^{-4}(g^2+g'^2),\nonumber
    \\
    &c_{HD}=1.75\times 10^{-3},~c_{6}=3.76\times 10^{-5},~c_{4hWW}=1.09\times 10^{-3} g^2,~c_{4hZZ}=5.47\times 10^{-4} (g^2+g'^2).
\end{align}

\section{Background dependent trilinear and quartic Higgs self-couplings}
\begin{align}
    \lambda_3(\phi_b)=&\frac{\phi _b M_h^2 e^{-\frac{M_h^2 \left(\phi _b^4-2 v^2 \phi _b^2\right)}{8 \Lambda ^4 v^2}} \left(\phi _b^2 M_h^4 \left(\phi _b^2-v^2\right){}^3
   \left(e^{\frac{M_h^2 \left(\phi _b^4-2 v^2 \phi _b^2\right)}{4 \Lambda ^4 v^2}}+1\right)+24 \Lambda ^8 v^4 \left(e^{\frac{M_h^2
   \left(\phi _b^4-2 v^2 \phi _b^2\right)}{4 \Lambda ^4 v^2}}+1\right)\right)}{16 \Lambda ^8 v^6 \cosh ^{\frac{3}{2}}\left(\frac{M_h^2 \left(\phi _b^4-2
   v^2 \phi _b^2\right)}{8 \Lambda ^4 v^2}\right)}\nonumber
   \\
   &+\frac{\phi _b M_h^2 e^{-\frac{M_h^2 \left(\phi _b^4-2 v^2 \phi _b^2\right)}{8 \Lambda ^4 v^2}} \left(6 \Lambda ^4 v^2 M_h^2 \left(-4 v^2 \phi _b^2+3 \phi
   _b^4+v^4\right) \left(e^{\frac{M_h^2 \left(\phi _b^4-2 v^2 \phi _b^2\right)}{4 \Lambda ^4 v^2}}-1\right)\right)}{16 \Lambda ^8 v^6 \cosh ^{\frac{3}{2}}\left(\frac{M_h^2 \left(\phi _b^4-2
   v^2 \phi _b^2\right)}{8 \Lambda ^4 v^2}\right)}
   \end{align}
   \begin{align}
   \lambda_4(\phi_b)=&\frac{M_h^2 e^{\frac{M_h^2 \left(\phi _b^4-2 v^2 \phi _b^2\right)}{8 \Lambda ^4 v^2}} \left(M_h^6 \left(\phi _b^3-v^2 \phi _b\right){}^4
   \left(e^{\frac{M_h^2 \left(\phi _b^4-2 v^2 \phi _b^2\right)}{4 \Lambda ^4 v^2}}-1\right)+48 \Lambda ^{12} v^6 \left(e^{\frac{M_h^2 \left(\phi _b^4-2
   v^2 \phi _b^2\right)}{4 \Lambda ^4 v^2}}+1\right)\right)}{8 \Lambda ^{12} v^8
   \left(e^{\frac{M_h^2 \left(\phi _b^4-2 v^2 \phi _b^2\right)}{4 \Lambda ^4 v^2}}+1\right){}^2}\nonumber
   \\
   +&\frac{M_h^2 e^{\frac{M_h^2 \left(\phi _b^4-2 v^2 \phi _b^2\right)}{8 \Lambda ^4 v^2}} \left(12 \Lambda ^8 v^4 M_h^2 \left(-14 v^2 \phi _b^2+17 \phi _b^4+v^4\right) \left(e^{\frac{M_h^2
   \left(\phi _b^4-2 v^2 \phi _b^2\right)}{4 \Lambda ^4 v^2}}-1\right)\right)}{8 \Lambda ^{12} v^8
   \left(e^{\frac{M_h^2 \left(\phi _b^4-2 v^2 \phi _b^2\right)}{4 \Lambda ^4 v^2}}+1\right){}^2}\nonumber
   \\
   -&\frac{M_h^2 e^{\frac{M_h^2 \left(\phi _b^4-2 v^2 \phi _b^2\right)}{8 \Lambda ^4 v^2}} \left(12 \Lambda ^4 \phi _b^2 M_h^4 \left(v^2-3 \phi _b^2\right) \left(v^3-v \phi
   _b^2\right){}^2 \left(e^{\frac{M_h^2 \left(\phi _b^4-2 v^2 \phi _b^2\right)}{4 \Lambda ^4 v^2}}+1\right)\right)}{8 \Lambda ^{12} v^8
   \left(e^{\frac{M_h^2 \left(\phi _b^4-2 v^2 \phi _b^2\right)}{4 \Lambda ^4 v^2}}+1\right){}^2}
\end{align}
\bibliographystyle{apsrev4-1}
\bibliography{mybib.bib}
\end{document}